\documentclass[showpacs,prb]{revtex4}

\usepackage{graphicx,amsfonts,amssymb}

\begin{document}

\title{New two dimensional $S = {1 \over 2}$ Heisenberg
antiferromagnets: synthesis, structure and magnetic properties}

\author{F.M. Woodward, A.S. Albrecht, C.M. Wynn, C.P. Landee}
\affiliation{Department of Physics, Clark University\\
Worcester, MA 01610}
\author{M.M. Turnbull}
\affiliation{Carlson School of Chemistry and Biochemistry, Clark University\\
Worcester, MA 01610}
\date{\today}
\begin{abstract}
The magnetic susceptibility and magnetization of two new layered  S=$1\over{2}$ 
Heisenberg antiferromagnets with moderate exchange are reported. 
The two isostructural compounds,  (2-amino-5-chloropyridinium)$_2$CuBr$_4$ 
((5CAP)$_2$CuBr$_4$) and (2-amino-5-methylpyridinium)$_2$CuBr$_4$ 
((5MAP)$_2$CuBr$_4$), contain S=${1\over2}$, Cu(II) ions related by C-centering, 
yielding four equivalent nearest neighbors.  The crystal structure of 
newly synthesized compound, (5CAP)$_2$CuBr$_4$, shows the existence of layers of 
distorted copper(II)-bromide tetrahedra parallel to the $ab$ plane, separated by the 
organic cations along the $c$ axis.  Magnetic pathways are available through 
the bromide-bromide contacts within the layers and provide for moderate 
antiferromagnetic exchange. Susceptibility measurements indicate interaction 
strengths to be 8.5(1) K, and 6.5(1) K  and ordering temperatures of 5.1(2) K and 3.8(2) K
for (5CAP)$_2$CuBr$_4$ and  (5MAP)$_2$CuBr$_4$ respectively. High field magnetization 
experiments on both compounds show upward curvature of M(H,T).  Magnetization measurements 
made at T = 1.3 K show saturation occurs in (5MAP)$_2$CuBr$_4$ at 18.8 T 
and in (5CAP)$_2$CuBr$_4$ at 24.1 T.  The magnetization curves are 
consistent with recent theoretical predictions.  Single crystal magnetization
measured at 2.0 K indicate a spin flop transition at 0.38 T and 0.63 T for
(5CAP)$_2$CuBr$_4$ and (5MAP)$_2$CuBr$_4$ respectively.
\end{abstract}
\maketitle

\section{Introduction}

\par{For over two and a half decades, low-dimensional magnetism has played an 
integral role in the understanding of phase transitions and critical 
phenomena.  During the past dozen years, there has been enhanced interest in 
low-dimensional magnetism in the condensed matter physics community due to the 
discovery of the copper-oxide superconductors which contain layers of 
S=$1\over{2}$, Cu(II) ions.  Experimental investigations of the insulating 
parent compounds of the superconductors, such as La$_{2}$CuO$_{4}$, have 
demonstrated the existence of strong antiferromagnetic intraplanar interactions 
($J \approx 1000$ K), with very weak interactions in the third dimension.
\cite{Birgeneau90}  It has been proposed that the formation of Cooper pairs 
in the the doped systems can be understood in terms of the antiferromagnetic 
interactions within the layers.\cite{Sokol93}  The consequent desire to 
understand the magnetic properties of the two-dimensional (2D) S=${1\over 2}$ 
(or Quantum) Heisenberg antiferromagnet (2D QHAF) has led to a large amount of
theoretical \cite{Manousakis91} and experimental research.}

\par{The current physical realizations of the 2D QHAF are inappropriate for 
examination of a number of important theoretical predictions, particularly 
those involving field-dependent properties and higher relative temperatures,
T/J $\geq$ 1. The large exchange strengths in the copper oxides will require 
magnetic fields exceeding 1,500 Tesla to bring the magnetic moments of the 
copper ions to saturation. For this reason, no theoretical studies of the 
in-field properties of the 2D QHAF appeared until very recently.
\cite{Yang97,Zhitomirsky98,Zhitomirsky99}  The few previously known non-oxide 
examples of 2D QHAF are characterized by much smaller exchange interactions 
but still have their own sets of limitations. Until more appropriate materials 
are available, a deeper understanding of the nature and properties of the 2D 
QHAF must be postponed.}

\par{Our research group endeavors to expand the available catalog of 
low-dimensional quantum antiferromagnets through the application of the 
principles of molecular-based magnetism.\cite{Kahn93}  We report here on a 
new family of 2D QHAF with relatively small intralayer exchange constants, 
permitting high-field studies. In this paper we present the first experimental 
investigations of low temperature magnetization curves for 2D QHAFs.  We report 
the synthesis and structure of (5CAP)$_2$CuBr$_4$ 
(5CAP = 2-amino-5-chloropyridinium) and the magnetic properties of both 
(5CAP)$_2$CuBr$_4$ and (5MAP)$_2$CuBr$_4$ (5MAP = 2-amino-5-methylpyridinium), 
two members of a family of insulating 2D S = ${1\over 2}$ Heisenberg 
antiferromagnets.  This family has the general chemical formula A$_2$CuX$_4$, 
where A =5CAP, or 5MAP and X = Br or Cl.  The copper ion is in a 2$^+$ 
oxidation state with a d$^9$ electron configuration, producing one unpaired 
spin (S = ${1\over 2}$) and nearly quenched orbital angular momentum 
($<$g$>$ $\approx$ 2.1).  The (5MAP)$_2$CuX$_4$ compounds were the first to be 
synthesized in this family.\cite{Place87,Zhou91} The (5CAP)$_2$CuX$_4$ 
compounds have been synthesized with the goals of reducing the number of 
protons in the cation for neutron scattering experiments and with the intent 
of increasing the interaction strength. Expressing the Hamiltonian as} 

\begin{equation}\label{ham}
H = J\sum_{<ij>}\vec{S}_{i}\cdot\vec{S}_j,
\end{equation}

\noindent{these materials have been found to have exchange strengths between 6
and 10 K, making it convenient to investigate their properties over a broad 
range of relative temperatures and applied fields. They are also easily 
prepared as single crystals.}

\section{Experimental}
\subsection{Synthesis and Characterization}
\par{Crystals of (5CAP)$_{2}$CuBr$_4$ were prepared by slow evaporation of an 
aqueous solution of anhydrous copper(II) bromide (2.23 g, 10 mmol), dilute 
(20\%) hydrobromic acid (8.1 g, 20 mmol) and 2-amino-5-chloropyridine 
(2.57 g, 20 mmol).  The reaction is shown below.}
\\
\begin{center}
\setlength{\unitlength}{1.5mm}
\begin{picture}(70,30)
\thinlines
\put(-1,15){2}
\put(1,7){\line(0,1){16}}
\put(26,7){\line(0,1){16}}
\put(1,7){\line(1,0){3}}
\put(1,23){\line(1,0){3}}
\put(26,7){\line(-1,0){3}}
\put(26,23){\line(-1,0){3}}
\put(10,12){\line(0,1){6}}
\put(20,12){\line(0,1){6}}
\put(10,18){\line(5,2){5}}
\put(15,20){\line(5,-2){5}}
\put(20,12){\line(-5,-2){4}}
\put(10,12){\line(5,-2){4}}
\put(15,15){\circle{6}}
\put(14,9){N}
\put(10,12){\line(-1,-1){3}}
\put(2,8){H$_2$N}
\put(20,18){\line(1,1){3}}
\put(23,21){Cl}
\put(27,14){
+ 2HBr + CuBr$_{2}$
$\stackrel{\textstyle{H_{2}O}}{\longrightarrow}$(5CAP)$_{2}$CuBr$_4$
}
\end{picture}
\end{center}
\noindent{Although no attempt has been made to maximize yield, the net mass of 
harvested crystals is typically 50\%-70\% of the theoretical yield.  Crystals 
as large as 650 mg have been grown.  The crystals are a very deep maroon color, 
and for sizes larger than a few milligrams, appear black.  Combustion analysis 
agrees with theoretical calculations.  Analysis for 
(C$_{5}$H$_{6}$N$_{2}$Cl)$_{2}$CuBr$_{4}$, calculated(\%): C, 18.73; N, 8.74; 
H, 1.89, found(\%): C, 18.64; N, 8.47; H, 1.82.  The infrared spectrum has the 
following principle bands: IR(KBr) $\nu$(cm$^{-1}$): 3424 m, 3307 m, 1662 s, 
1608 s, 1331 m, 820 m, 661 m.  The letters s and m indicate strong and medium 
intensity.  Crystals of (5MAP)$_{2}$CuBr$_4$ were prepared by the method 
described above with the substitution of 2-amino-5-methylpyridine
(2.22 g, 20 mmol) for the 2-amino-5-chloropyridine.}

\subsection{X-ray data collection}

\par{The X-ray diffraction data for (5CAP)$_{2}$CuBr$_4$ were collected at 
-130 $^{\circ}$C using a Siemens P4 diffractometer.  The crystal data and 
structure refinement parameters are shown in Table \ref{crys}.  Optimization 
of the orientation matrix and lattice parameters was done using least-squares 
calculation on 16 reflections in the range $4.64^{\circ}<\theta<12.23^{\circ}$.
Standard reflections (3) were monitored every 97 reflections to measure
variations.  The standard reflections varied by only 7.6\%.  A total of 3173 
reflections were measured using an $\omega$ scan. Upon data reduction, 1598 
unique reflections remained with 1181 having the criterion $|F|>2\sigma$.
Details of the crystal structure and data collection method of 
(5MAP)$_{2}$CuBr$_4$ are given in the work of Place and Willett.\cite{Place87}
}

\subsection{Magnetic measurements}
\par{The measurements of susceptibility and low field magnetization for single 
crystal samples of (5CAP)$_{2}$CuBr$_4$ and (5MAP)$_{2}$CuBr$_4$ were made 
using a Quantum Design SQUID magnetometer.  Initial single crystal studies were 
hampered by the single crystal samples shattering as a result of thermal cycling.  
This problem was overcome by embedding them in Emerson and Cummings Stycast 1266 
epoxy.  The orientation of the crystals was established by correlation of crystal
morphology to X-ray structure.  This correlation was verified by room temperature
EPR on several single crystal samples.  The determination of the relation between the 
magnetic axes relative to crystal morphology was accomplished by observing the g-values 
as a function of angle for three orthogonal rotation of the crystal.  The powder and single 
crystal susceptibility data for the two compounds were measured in fields up to 3 T using 
a Quantum Design SQUID. Corrections have been made for temperature independent 
paramagnetism (TIP = 60x10$^{-6}$ $cm^2\over{mol}$) and the intrinsic 
diamagnetism (DIA= -329x10$^{-6}$ $cm^2\over{mol}$ for 5MAP and 
-280x10$^{-6}$ $cm^2\over{mol}$ for 5CAP) of the samples.  High field 
magnetization data were collected for powder samples using a VSM at the 
National High Field Magnet Laboratory in Tallahassee, Florida. Fields up to 30 
Tesla were applied to the samples at various temperatures.  The EPR data,
including single crystal alignment, were collected on a Bruker EMX spectrometer 
operating at 9.3 GHz. Low temperature EPR data were collected using an Oxford 
ESR-910 helium flow cryostat.}

\section{Results}

\subsection{Crystal structure}
\par
\begin{figure}[h]
{\includegraphics{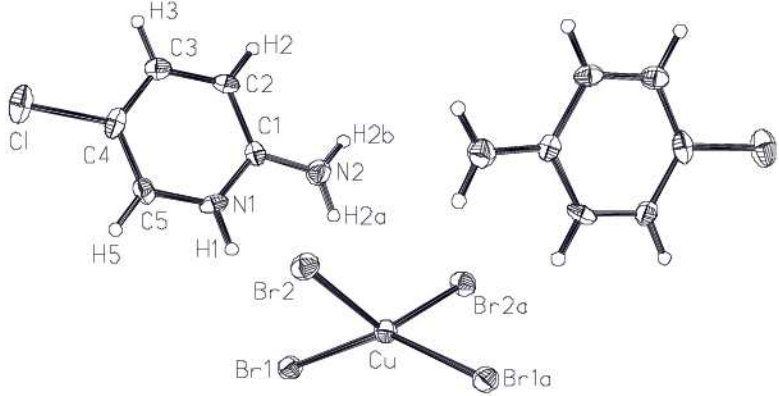}}
\caption{Molecular unit of (5CAP)$_2$CuBr$_4$}
\label{molunit}
\end{figure}

\begin{figure}[h]
{\includegraphics{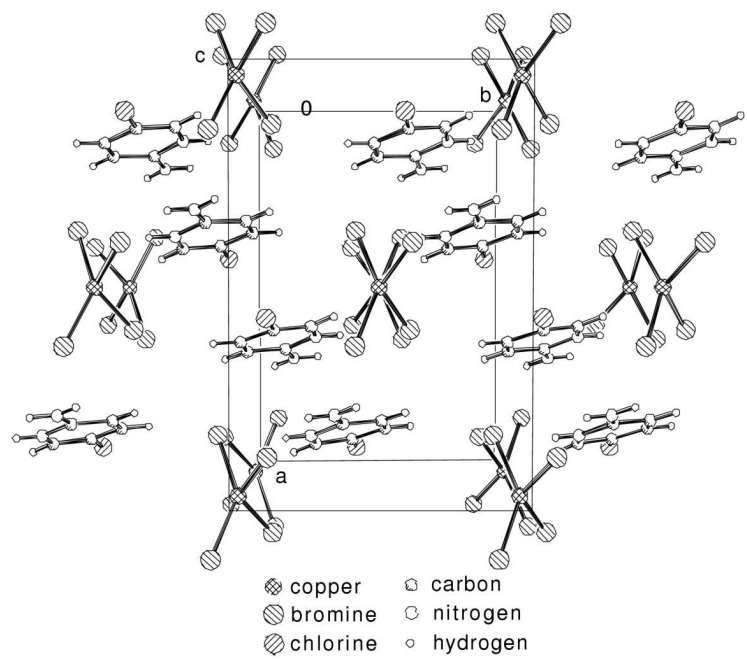}}
\caption{View down $c$-axis of (5CAP)$_2$CuBr$_4$ showing two adjacent 
C-centered CuBr$_{4}^{2-}$ planes in their eclipsed configuration.}
\label{struc_top}
\end{figure}

{Crystals of (5CAP)$_{2}$CuBr$_4$ are monoclinic in the space group C2/c, 
with $a$ = {13.050(5) \AA}, $b$ = {8.769(3) \AA}, $c$ = {15.810(5) \AA}, and 
$\beta = 94.31(3)^\circ$.  The atomic coordinates and equivalent isotropic 
displacement parameters are given in Table \ref{table2}.  Selected bond 
distances and angles are presented in Table \ref{table3}.  The structure of 
the molecular unit is shown in Figure \ref{molunit}.  Within the unit cell, 
the copper tetrabromide dianions sit at the edges and centers of planes 
parallel to the $ab$ plane ($c$ = 0.25, 0.75), related by unit cell 
translations and C-centering, respectively (Fig. \ref{struc_top}).  These 
copper tetrahedra are flattened with the mean Br--Cu--Br large angle 
$\overline{\theta} \approx 137^{\circ}$.  The copper ions lie on the two-fold 
symmetry axes.  Consequently, each tetrahedra has its compression axis 
parallel to the $b$-axis, eliminating any canting of the local g-tensor.  
Equivalent layers of CuBr$_{4}^{2-}$ tetrahedra are located one-half unit cell 
apart along the c-axis. Each copper site is related to one in the adjacent 
layers by the $c$-glide symmetry operation.}
	
\par{The copper tetrahedra are tightly packed along the diagonals of the $ab$ 
layers, with the separation between nearest neighbor copper(II) ions in
this direction being {7.86 \AA}. Such pairs of copper atoms are related
by the C-centering operation. The Br$\cdots$Br separation between 
adjacent tetrahedra along the diagonal is only {4.35 \AA}, approximately twice 
the radius of the bromide ion. The dihedral angle formed by the 
Cu--Br$\cdots$Br--Cu pathway is approximately 22$^{\circ}$. Such halide-halide 
contacts are known to create weak antiferromagnetic interactions
\cite{Zhou91,Willett88} which decrease rapidly with increased 
Br$\cdots$Br separation. The Br$\cdots$Br contact distances along 
the $a$ and $b$ axes are more than {10 \AA} and {7 \AA} respectively, so the 
intralayer magnetic interactions must take place between copper ions related 
by C-centering. Since each copper ion has four such identical neighbors, this 
lattice is $magnetically$ equivalent to a square 2D lattice.}

\par
\begin{figure}[h]
{\includegraphics{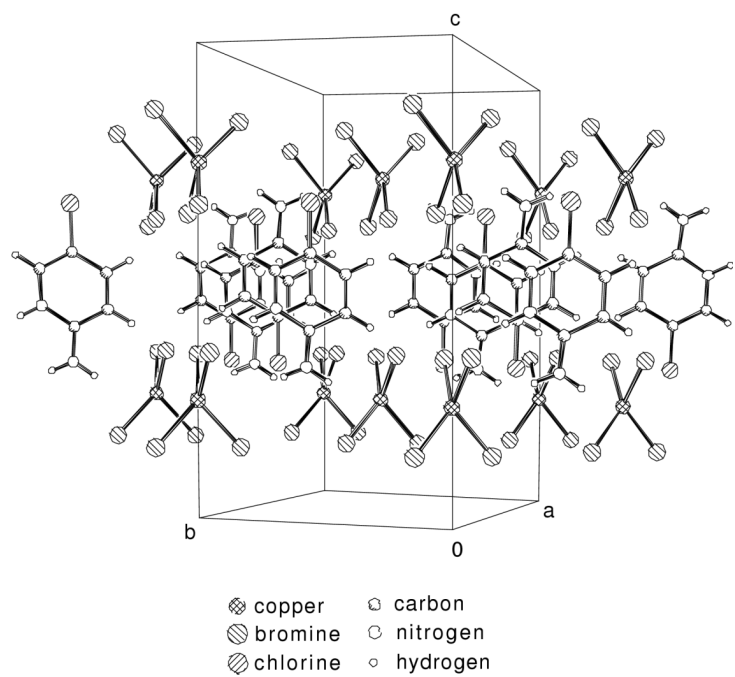}}
\caption{View down the $a$-axis of (5CAP)$_2$CuBr$_4$ showing the cross 
section of the planes and the orientation of the organic groups. The dashed 
lines mark the two sets of interplanar $Br\cdots$Br contacts which cause the 
interplanar exchange J|prime.}
\label{struc_cross}
\end{figure}

{The layers of copper bromide tetrahedra are stabilized into a 3D array by 
the organic cations which lie between the CuBr$_{4}^{2-}$ layers.  They are 
stacked parallel to the $ab$ diagonal and separated by {3.4 \AA}.  Successive 
pyridinium rings within the stack are related by a two-fold rotation.  Looking 
down the stacking axis, the pyridinium substituent which points up along the 
$c$-axis alternates between the 2-amino and the 5-chloro (see Figure 
\ref{struc_cross}).  The planes of the pyridine rings are tilted approximately 
70$^{\circ}$ with respect to the copper planes, resulting in a 
separation of copper centers in neighboring planes of {7.88 \AA}.  Weak 
hydrogen bonding between the pyridinium hydrogen (H1) and Br1 (refer to 
Fig. \ref{molunit}, $d_{H1--Br1} = 3.32$ {\AA}) helps stabilize the structure.  
Very weak hydrogen bonding may also occur between the amino hydrogens 
(H2a, H2b) and two neighboring bromines from different tetrahedra 
($d_{N2--Br2,Br2a}$ = 3.51, {3.61 \AA}).}

\par{The magnetic layers are coupled in the third dimension by an interlayer 
interaction J' that occurs through Br$\cdots$Br contacts along the 
$c$-axis (Fig. \ref{struc_cross}).  Copper sites along the $c$-axis are related 
by two identical Br$\cdots$Br contacts at a distance of {4.83 \AA}, with a 
dihedral angle of approximately 21$^{\circ}$.  The extra {0.48 \AA} separation 
in  the intralayer Br$\cdots$Br contact distances will lead to a 
significant reduction in the J'/J ratio.}

\par{(5MAP)$_2$CuBr$_4$\cite{Place87} is isostructural with (5CAP)$_2$CuBr$_4$.
The room temperature lattice parameters for (5MAP)$_2$CuBr$_4$ are $a$ = 
13.715(2) \AA, $b$ = 8.7162(2) \AA, $c$ = 16.013(4) \AA, and $\beta = 
93.79(2)^\circ$, reflecting its slightly larger unit cell.  The intraplanar 
Br$\cdots$Br distance is {4.54 \AA}, which is significantly longer than the 
corresponding value in (5CAP)$_2$CuBr$_4$ of {4.35 \AA}.  The separation between 
the layers is also significantly enhanced due to the bulk of the methyl 
substituent resulting in a separation of {4.97 \AA}.} 

\subsection{Powder susceptibility}
\par
\begin{figure}[h]
{\includegraphics[width=2.5in,angle=-90]{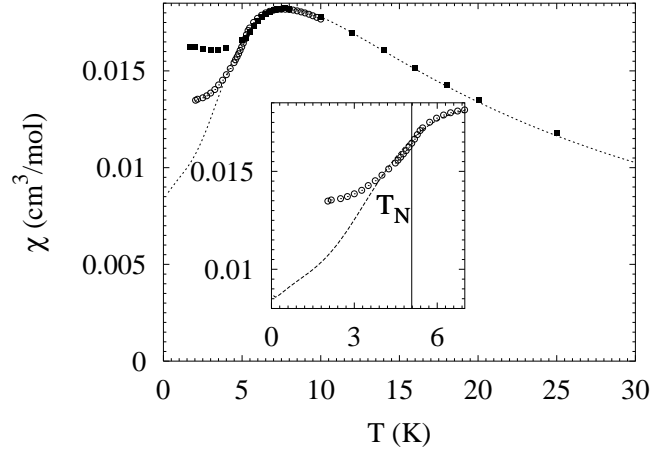}}
\caption{$\chi_m$ vs T  for (5CAP)$_2$CuBr$_4$.  The dashed line is
the 2D QHAF model using the parameters J = 8.5(2) K and g = 2.11(2). The data, {\large $\circ$},
were collected at H = 0.1 T and the $\blacksquare$ were collected at H = 1.0 T. The vertical line 
in the insert marks the ordering temperature of 5.08 K as determined by specific heat
measurements\cite{Sorai00}. }

\label{cap_chi}
\end{figure}

{The molar magnetic susceptibility ($\chi$$_m$) as a function of 
temperature for a powder of (5CAP)$_2$CuBr$_4$ is shown in Figure 
\ref{cap_chi}. A broad maximum is observed with the maximum value in 
$\chi$$_m$ (18.3 x 10$^{-3}$ ${cm^{3}}\over{mol}$) occurring near 8.0 K.  The 
data have been compared to the theoretical predictions and simulation for the susceptibility 
of the 2D QHAF (described in the Discussion). The dashed line shown in Figure 
\ref{cap_chi} represents a curve fit to the data resulting in an exchange 
interaction strength J = 8.5(1) K and g$_{ave}$ = 2.11(2). This value of 
g$_{ave}$ is in good agreement with powder and single crystal room temperature
EPR measurements.  The magnetic susceptibility fitting procedure included only 
data at temperatures greater than 5.2 K, since the specific heat studies\cite{Sorai96} of 
(5CAP)$_2$CuBr$_4$ show the existence of an ordering 
transition at T$_{N}$ = 5.08 K.  The dashed line shows the model expression 
for the ideal 2D QHAF with the same parameters extended down to T = 0. The low field 
powder susceptibility shows no anomaly at the ordering transition, but does 
break away from the model curve at a temperature very close to T$_N$.  The data collected 
in a field above the spin-flop transition (Sec. E below), shows a much stronger
deviation from the model curve at T$_N$.}  

\par{The data for (5MAP)$_2$CuBr$_4$ are shown in Figure \ref{map_chi}.  The 
susceptibility of (5MAP)$_2$CuBr$_4$ is qualitatively identical to that of 
(5CAP)$_2$CuBr$_4$, with a slightly lower temperature for the maximum 
susceptibility ($\approx$ 6 K).  Comparison of these data to the model curve 
yields an interaction strength of J = 6.5(1) K and g$_{ave}$ = 2.07(2).}

\subsection{High field magnetization}
\par{The magnetizations as a function of field at T = 1.3 K for 
(5CAP)$_2$CuBr$_4$ and (5MAP)$_2$CuBr$_4$are shown in Figure 
\ref{mvh}(a) plotted on a normalized scale M/M$_{sat}$ where M$_{sat}$ were 
determined to be 5980 and 5880 emu/mol respectively.  To our knowledge, this is the first 
report of the full magnetization curve for any 2D QHAF. (A preliminary report has appeared 
elsewhere\cite{Turnbull99}). Note the upward curvature present in both data 
sets.  The saturation fields appear to be close to 19 T and 24 T, respectively.
Although these estimates are crude, we do note that the ratio of saturation 
fields (19 T/24 T = 0.79) is quite close to the ratio of exchange strengths as 
determined by the susceptibility data (6.8 K/8.5 K = 0.80).}
 
\par{In Figure \ref{mvh}(a), the data are plotted again on a normalized 
scale, M/M$_{sat}$ versus H/H$_{sat}$ where H$_{sat}$ is 18.8 T and 24.1 T 
for (5MAP)$_2$CuBr$_4$ and (5CAP)$_2$CuBr$_4$ respectively.  These values of H$_{sat}$ 
were determined from mean field approximations using the interaction strengths,
J, as determined from the powder susceptibility data for each compound.  
Details for this procedure are described in the Discussion section.  Included 
in Figure \ref{mvh} are results from numerical calculations of M(H, T = 0) 
for finite 1D\cite{Griffiths64},  2D\cite{Yang97,Zhitomirsky98} Heisenberg 
systems, and Monte-Carlo simulation at T/J = 0.2 \cite{troyer2}.  The 
magnetization data of (5MAP)$_2$CuBr$_4$ is slightly higher than that of 
(5CAP)$_2$CuBr$_4$ until a point just below the saturation field.}

\par{The temperature dependence of the molar magnetization as a function of 
field for a powder sample of (5MAP)$_2$CuBr$_4$ is shown in Figure 
\ref{map_mvh}. The data were collected at three different temperatures 
(T = 1.3, 2.4, 4.4 K) corresponding to relative temperatures T/$J$ of 0.19, 
0.35, and 0.65 respectively.  The data at 1.3 K and 2.4 K both exhibit upward 
curvature and saturate at approximately 20 T. The upward curvature is no 
longer present in the higher temperature data.}
 
\subsection{Single Crystal Susceptibility}
\par{Figure \ref{cxtal}(a) shows the single crystal magnetic susceptibility 
data for (5CAP)$_2$CuBr$_4$ collected with the applied magnetic field both in and 
perpendicular to the 2D magnetic layer.  The powder data as well as the 2D QHAF 
model curve are also included in this figure.  Notice that at 5.1 K the data 
sets diverge from one another.  This is an expected result for an isotropic 
2D QHAF as it goes through dimensional crossover to an anisotropic 3D ordered 
state. In a 3D QHAF with an Ising anisotropy, the three distinct susceptibility curves 
are $\chi$$_\|$, dropping toward zero, $\chi$$_{\perp}$, staying at or close 
to $\chi$$_{max}$, and the powder data, $\chi$$_{pow}$, which should fall in between 
the two previous curves (2/3 $\chi$$_{max}$).  Note that the onset of the 
three dimensional behavior in the single crystal data begins at a temperature 
(5.1 K) which is in excellent agreement with the critical temperature 
determined by specific heat studies\cite{Sorai96}. The (5MAP)$_2$CuBr$_4$ data 
shown in Figure \ref{cxtal}(b) exhibit the same dimensional crossover 
behavior.  For the (5MAP)$_2$CuBr$_4$ data the onset of 3D order occurs at a 
lower temperature of 3.8 K.  The single crystal data for both compounds were 
collected with H$_{applied}$ = 0.2 T and include corrections for diamagnetic and temperature 
independent paramagnetic contribution.  Note that the data in Figures
\ref{cxtal}(a) and (b) have been scaled by the Curie constant for a spin 1/2 system
, C = 0.375(g/2)$^2$, where the g-value is g$_a$, g$_b$=g$_c$, or g$_{pow}$ depending
upon the data set.}

\subsection{Single Crystal Magnetization}

\par{Figure \ref{mxtal}(b) shows single crystal magnetization for 
(5MAP)$_2$CuBr$_4$ at T = 2.1 K with magnetic field applied in three orthogonal 
directions. Clearly visible in Figure \ref{mxtal}(b) is a change in slope of the 
magnetization data when the magnetic field is applied parallel to the 
crystallographic a-axis.  The change of slope in the magnetization data occurs 
at H$_{applied}$ = 0.63 T.  The inflection is only observed when the 
field is applied along the a-axis and is not found when the field is applied
along the {\it b} or {\it c*} as seen in the other two data sets displayed in figure
\ref{mxtal}(b).  The (5CAP)$_2$CuBr$_4$ single crystal magnetization curve also at
T = 2.1 K, in Figure \ref{mxtal}(b), also shows a change in slope at 
H$_{applied}$ = 0.30 T when H$_{applied}$ is along the a-axis.  In the case of 
(5CAP)$_2$CuBr$_4$ data shown in Figure \ref{mxtal}(a), the transition is less pronounced 
and even appears in one of the perpendicular curves.  This is due to difficulties of 
aligning the (5CAP)$_2$CuBr$_4$ crystal in the applied field.  Slight misalignment results 
in a mixing of the features of orthogonal magnetization curves.}

\subsection{EPR}
\par{The angular dependence of the single crystal data for three orthogonal rotations
in the {\it ab}, {\it bc*}, and {\it ac*} planes respectively are found in Figure 
\ref{eprsc}.  In the case of (5CAP)$_2$CuBr$_4$,Figure \ref{eprsc}(a), the angular study clearly 
shows two principle g values: g$_b$=2.22 and g$_a$=g$_{c*}$=2.06.  Powder simulation for 
(5CAP)$_2$CuBr$_4$ yields two g values, g$_\perp$ = 2.22 and g$_{||}$ = 2.06.  The 
Jahn-Teller compression of the copper's tetrahedral environment along the $b$-axis is the 
cause of the 7.2\% difference between g-values along the different axes.  The same is true 
for (5MAP)$_2$CuBr$_4$, as seen in Figure \ref{eprsc}(b).  Here g$_b$=2.24 while 
g$_a$=g$_{c*}$=2.05.  This is in agreement with the powder simulation values, g$_\perp$ = 
2.23 and g$_{||}$ = 2.06, yielding difference of 7.6\% for this compound.}

\par{The room temperature X-band powder EPR spectra display evidence of a slightly 
anisotropic copper signal for each compound.  For (5MAP)$_2$CuBr$_4$5, the single 
crystal average g-value is $<g_{sc}>$ = $\sqrt{{1\over{3}}(g_x^2+g_y^2+g_z^2)}$ = 
2.12.  The average g-value as determined by a comparison of powder EPR data to powder 
simulation, using Bruker's EPR simulation package SimFona, is $<g_{sim}>$ = 2.12.  
The corresponding (5CAP)$_2$CuBr$_4$ g-values are $<g_{sc}>$ = 2.11 and $<g_{sim}>$ 
= 2.11\cite{Poole72}.}

\par{The low temperature signals for X-band are quite remarkable.  Figure
\ref{eprlt} shows room temperature powder data as the top spectra and 
low temperature spectra, T=3.0 K as in the bottom half of the figure.  The 
(5CAP)$_2$CuBr$_4$ data are on the left and (5MAP)$_2$CuBr$_4$ data are on 
the right.  The rich and complex spectra observed in the X-band data in the
lower half of Figure \ref{eprlt} are not found in the Q-band spectra.} 

\section{Discussion}
\subsection{Structure}
\par{The crystal structures of the compounds in the (5CAP/5MAP)$_2$CuBr$_4$ 
family show that the copper ions lie on a C-centered lattice, with four 
equivalent nearest neighbors parallel to the $ab$ plane (Fig. \ref{struc_top}
).  The Br$\cdots$Br contacts between CuBr$_{4}^{2-}$ tetrahedra along the
diagonals lead to antiferromagnetic interactions of a few Kelvin, consistent 
with the values observed for other structures in which copper tetrabromide 
anions are in contact.\cite{Zhou91,Willett88}  The distances to the next 
nearest neighbors within the planes are much greater, with negligible 
Br$\cdots$Br ion contacts. The magnetic interactions between next-nearest 
neighbors (J$_{nnn}$) can therefore be ignored. The equivalence of all nearest 
neighbor interactions, caused by the C-centering, plus the absence of 
J$_{nnn}$, permits the magnetic layers to be considered as a square magnetic 
lattice despite the absence of four-fold symmetry.} 
	
\par{The strength of the intraplanar interaction is predominantly dependent 
upon the value of the Br$\cdots$Br contacts.  This is evidenced by the change 
in interaction strength going from (5MAP)$_2$CuBr$_4$ to (5CAP)$_2$CuBr$_4$. 
The substituent in the five position on the pyridine ring protrudes into the 
copper tetrahedra layer (Figure \ref{struc_cross}).  The larger methyl group 
forces the tetrahedra farther apart, increasing the Br$\cdots$Br contact 
distance between adjacent tetrahedra from {4.35 \AA} for (5CAP)$_2$CuBr$_4$ to 
{4.54 \AA} for (5MAP)$_2$CuBr$_4$) and causing the decrease in the magnitude 
of the exchange interaction from 8.5 K for (5CAP)$_2$CuBr$_4$ to 6.8 K
for (5MAP)$_2$CuBr$_4$.  Such sensitivity of the structure to the 
size of the 5-substituent provides some adjustability in the magnetism of 
these systems, always a desirable goal of magneto-chemistry.}

\par{Studies have also been carried out on the chloride analogs to these 
systems, ((5CAP)$_2$CuCl$_4$ and (5MAP)$_2$CuCl$_4$).\cite{Hammar97}  The 
van der Waal's radius of a chloride ion is smaller than that of a bromide ion, 
whereas the unit cell constants of (5CAP)$_2$CuCl$_4$ and (5MAP)$_2$CuCl$_4$ 
are nearly the same as the bromide complexes.  Therefore, the absolute overlap 
of the Cl$\cdots$Cl wavefunctions between neighboring tetrahedra is 
considerably less than that of the Br$\cdots$Br overlap.  The smaller 
van der Waal's radius of the chloride ion produces a weaker exchange between 
the copper centers (1.14 K, 0.76 K respectively).}

\subsection{Susceptibility}
\par{The susceptibility of the 2D QHAF was originally calculated by Rushbrooke 
using high temperature series expansions (HTSE).\cite{Rushbrooke74} 
This procedure predicted a broad maximum in $\chi_m$ at $T_{max} \approx J$ 
but was invalid for temperatures below T$_{max}$. The discovery of high 
temperature superconductivity in 1986 stimulated renewed interest in the 
2D QHAF model and more extensive studies have been done. The techniques used
to evaluate the susceptibility include spin wave expansions\cite{Takahashi89} 
about T = 0, Pad$\acute{e}$ approximant extensions to the results of HTSE\cite{Singh90}, 
and quantum Monte Carlo calculations \cite{Makivic91,Troyer}. The low 
temperature susceptibility is predicted \cite{Takahashi89,Troyer} to approach the 
limiting value in a linear manner} 

\begin{equation}
{{\chi_m J}\over{C}} = {{\chi_0 J}\over{C}}+ {{0.15616T}\over{J}}
\end{equation}

\noindent{where $\chi_0$J/C = 0.174. There is no evidence of unusual quantum 
behavior in the susceptibility as T $\rightarrow$ 0, as has been recently 
demonstrated to be the case for the 1D QHAF \cite{Eggert94}. The maximum 
susceptibility has been shown to have a value of $\chi_{max}$ = 0.375(1) C/J 
at a temperature T$_{max}$ = 0.936 J\cite{Takahashi89}. The ratio of the two 
limiting susceptibilities $\chi_0/\chi_{max}$ = 0.416.}

\par{For the purposes of data analysis, the theoretical results 
were fit to an empirical expression for the susceptibility}
 
\begin{equation}
\chi_m = {{0.375g^2}\over{4T}}\sum_{n=1}^5{{{a_nK^n}\over{b_nK^n}}}
\end{equation}

\noindent{where K = J/T.  The coefficients a$_n$, b$_n$, listed in Table \ref{coef},  
were determined using a standard non-linear, least-squares fitting algorithm.  This 
functional form can be used to determine the J and g values from magnetic 
susceptibility data for any 2D QHAF.  This functional form accurately describes the
predicted susceptibility the 2D QHAF for a range of 0.15 $<$ T/J.}  

\par
\begin{figure}[h]
{\includegraphics[width=2.5in,angle=-90]{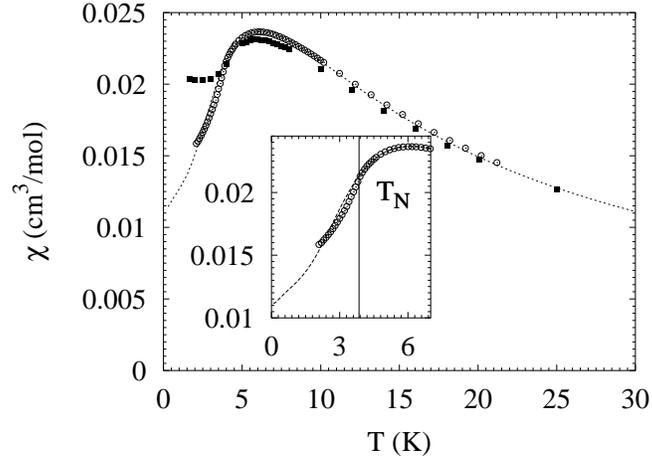}}
\caption{$\chi_m$ vs T  for (5MAP)$_2$CuBr$_4$.  The dashed line is
the 2D QHAF model using the parameters J = 6.5(2) K and g = 2.07(2). The data, {\large $\circ$},
were collected at H = 0.1 T and the $\blacksquare$ were collected at H = 1.0 T. The vertical line 
in the insert marks the ordering temperature of 3.8 K as determined by the deviation of
the powder susceptibility data from the ideal 2D QHAF curve.}
\label{map_chi}
\end{figure}

{The powder susceptibilities of (5CAP)$_2$CuBr$_4$, Figure \ref{cap_chi}, 
and (5MAP)$_2$CuBr$_4$, Figure \ref{map_chi}, are well described by this 
empirical expression for the 2D QHAF with the exchange strengths 8.5(1) K 
and 6.5(1) K, respectively. The best fits are shown as the dashed lines in Figures 
\ref{cap_chi} and \ref{map_chi}. The model curves based on these parameters (J and g) 
have been extended to zero temperature (dashed lines) and it is noted that the
experimental data break away from the theoretical curves at temperatures near 
5.1(2) K for (5CAP)$_2$CuBr$_4$ and 3.8(2) K for (5MAP)$_2$CuBr$_4$. Since the T$_N$ 
temperature of (5CAP)$_2$CuBr$_4$ is known to be 5.08 K by specific heat 
measurements\cite{Sorai96}, we can establish a close agreement between T$_N$ 
and the temperature of divergence between the experimental data and the 
theoretical prediction. On the basis of this comparison, we estimate the 
N$\acute{e}$el temperature of (5MAP)$_2$CuBr$_4$ to be 3.8(2) K.}

\par{Initially all of the powder susceptibility data were collected at 1 T, well above the
the field induced transition observed in the single crystal magnetization data.  This caused
the powder data to deviate sharply from the 2D QHAF curve below T$_N$, being dominated by
the $\chi_\perp$ behavior.  When collected in a field smaller than the spin-flop field, the data
still noticeably deviates from the model curve, but now approaches a value consistent with the 
mean field theory estimates of a 3D QHAF with a weak Ising anisotropy.  The powder 
susceptibilities in the ordered states approach constant values near 0.012 cm$^3$/mol and 
0.015 cm$^3$/mol for the (5CAP)$_2$CuBr$_4$ and (5MAP)$_2$CuBr$_4$ respectively.  These are near 
the theoretical limiting values of 2/3$\chi_\perp$, taken from the single crystal data, of 
0.011 cm$^3$/mol and 0.014 cm${^3}$/mol, respectively.}

\par{The single crystal susceptibilities for both of the compounds clearly show
dimensional crossover from isotropic 2D behavior to anisotropic 3D behavior by the 
appearance of $\chi$$_{\perp}$ and $\chi$$_{||}$ susceptibilities.  In (5CAP)$_2$CuBr$_4$, 
the onset of 3D order occurs at a temperature of 5.1 K as depicted in Figure 
\ref{cxtal}(a).  In the (5MAP)$_2$CuBr$_4$ the onset of 3D order begins
at 3.8 K and again is marked by the separation of the parallel and perpendicular
susceptibilities as shown in Figure \ref{cxtal}(b).}

\par{Analysis of the single crystal susceptibility data shows that there is an
internal anisotropy axis which lies along the crystallographic $a$-axis for each
compound.  The (5CAP)$_2$CuBr$_4$ and (5MAP)$_2$CuBr$_4$  $\chi_\|$ data shown in 
Figures \ref{cxtal}(a) and \ref{cxtal}(b) respectively, appear to fall well below the  
$\chi_\perp$ but do not quite extrapolate to $\chi_\|$(0) = 0 which is what is expected of a 
true 3D QHAF with an Ising anisotropy. The failure of the data to extrapolate to zero may
be due to either a misalignment of the sample in the applied field or to spin canting within
the ordered state.  From the extrapolated value of $\chi_\|$(0) the amount by which the sample 
is misaligned can be estimated.  The misalignment of the 5MAP sample would have to be about 
5$^\circ$ and in the 5CAP about 20$^\circ$. The method by which the samples were aligned in the
field allow for an error of at least 5$^\circ$, easily explaining the problems with the 
5MAP data.  An alignment error 20$^\circ$ in the 5CAP data is still not outside the realm of
possibility, when one considers that in the case of 5CAP the {\it a} and {\it b} axes do not 
lie directly along sample diagonals as was the case for 5MAP.  Also, there is an angle of 
5$^\circ$ between {\it c} and {\it c$^*$} and  4 $^\circ$ in the 5MAP.  While 
spin canting would explain a non-zero value of $\chi_\|$(0), structural considerations
tend to rule this out as an option.  Each copper atom is related to the next by the C-centering
in the lattice thus removing any possibility that any two copper sites would experience
different chemical environments.  It is possible that there is a structural phase
change as a function of temperature which would remove this symmetry, but no anomalies were
observed in the specific heat study\cite{Sorai96} of 5CAP except at T$_N$.  To conclusively 
determine whether spin canting exists in the 5CAP and 5MAP systems, a determination of the 
magnetic lattice by neutron scattering will have to be made.}

\par{Anisotropy in the CuBr$_4^{2-}$ complexes is due to distortions of the
coordination about the copper site from pure tetrahedral symmetry.  In 5CAP and 
5MAP, the distortions consist of a compression along the {\it b}-axis which
is reflected in the g-tensor anisotropy: g$_\|$ = g$_b$ and g$_\perp$ = 
g$_a$ = g$_c$.  Surprisingly, the unique magnetic axis, as determined by $\chi_\|$, is
not the same as the principle axis of the g-tensor.}

\par{Previous work by Willett\cite{willet_epr} has shown that for magnetic copper 
chloride compounds the easy axis lies along the unique axis of the coordination sphere
of the copper chloride tetrahedra, but in the case of the copper bromide compounds 
the easy ($\chi_\|$) axis tends to lie perpendicular to the compression axis.  Willett 
attributed this to a reversal in the sign of the spin orbit coupling parameters from the 
chloride to the bromide compounds.  Given that the {\it b}-axis is the bromide compression 
axis, the magnetic anisotropy axis is expected to lie along either the {\it a} or 
{\it c$^*$}.  It is not currently clear why a is selected over {\it c$^*$}, but the 
explanation may be found in the difference of the dipolar energies of the two 
configurations.}

\subsection{Magnetization}

\par
\begin{figure}[h]
{\includegraphics[width=2.5in,angle=-90]{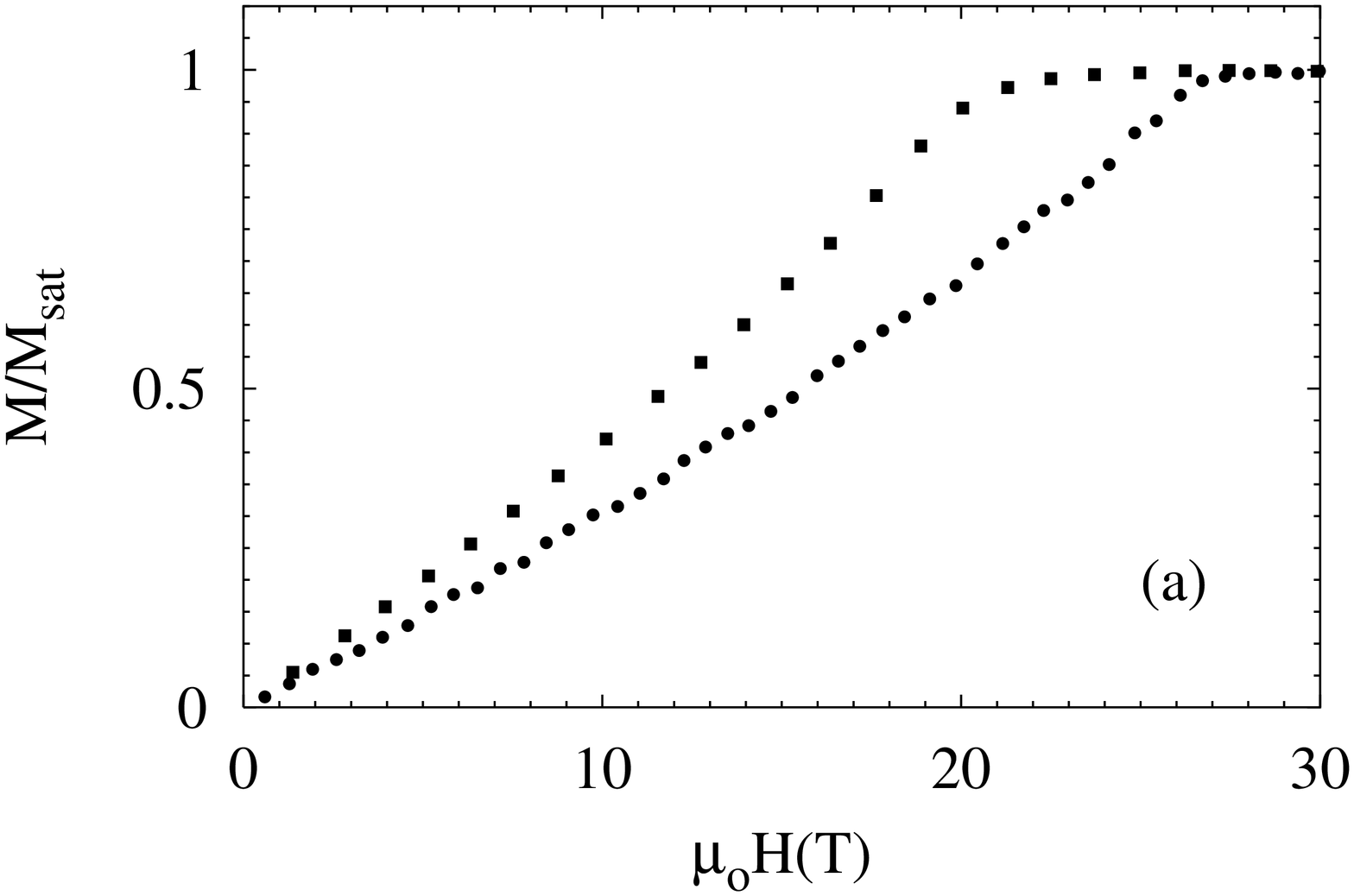}}
{\includegraphics[width=2.5in,angle=-90]{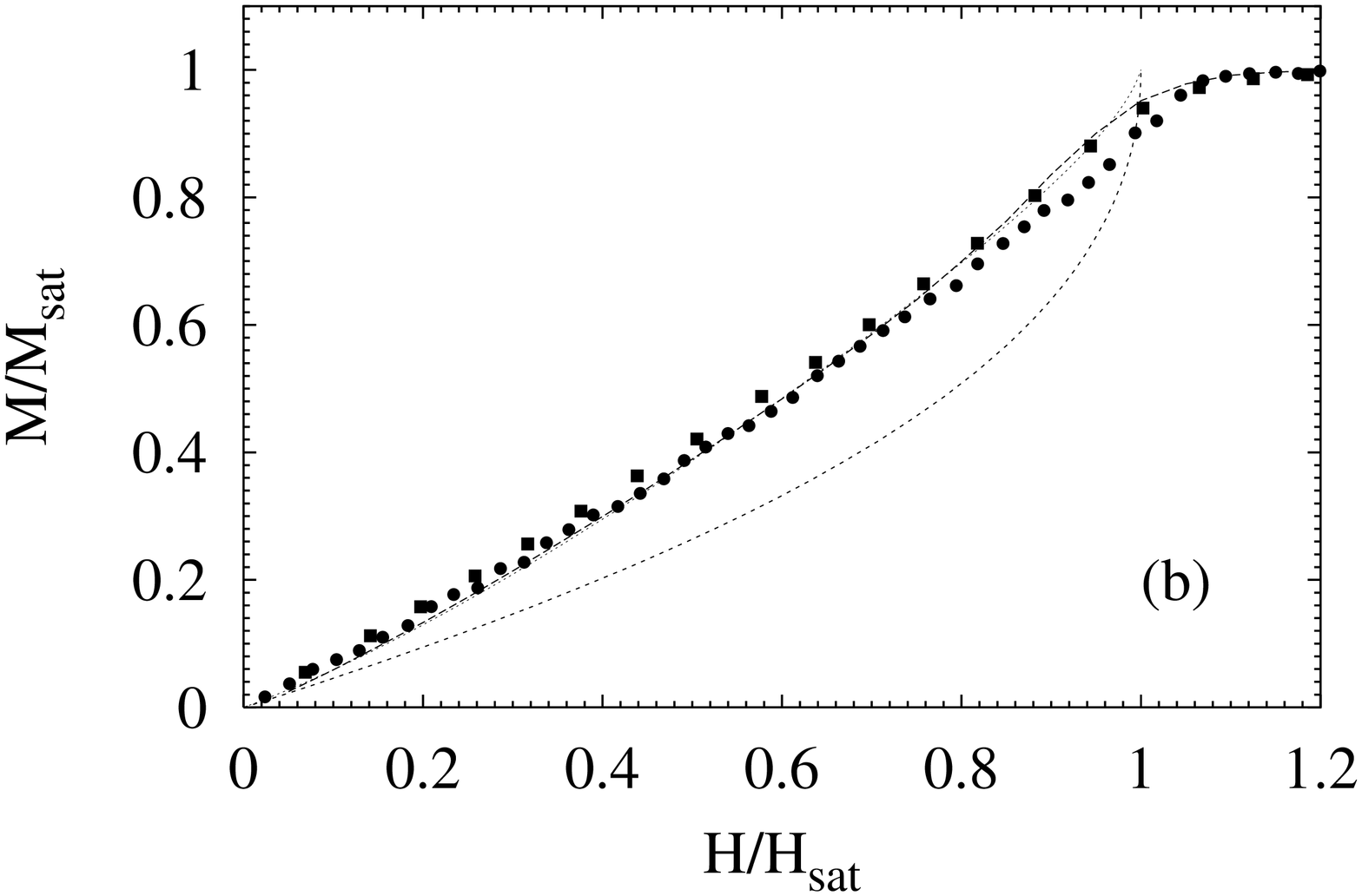}}
\caption{(a) Relative molar magnetization vs H (T)  at T = 1.3 K for powder samples of
(5MAP)$_2$CuBr$_4$: {\large$\bullet$}, (5CAP)$_2$CuBr$_4$: $\blacksquare$. 
(b) Same data as in (a) but plotted on a normalized field scale, H/H$_{sat}$ . 
H$_{sat}$ = 18.8 T for (5MAP)$_2$CuBr$_4$, H$_{sat}$ = 24.1 T for (5CAP)$_2$CuBr$_4$.  
The dotted line is a result of 2D numerical calculations at T = 0 K, the long dashed
line is the result of a Monte Carlo simulations at T/J = 0.2, and the dashed line 
is the result from 1d numerical calculations at T = 0 K.}
\label{mvh}
\end{figure}

\begin{figure}[h]
{\includegraphics[width=2.5in,angle=-90]{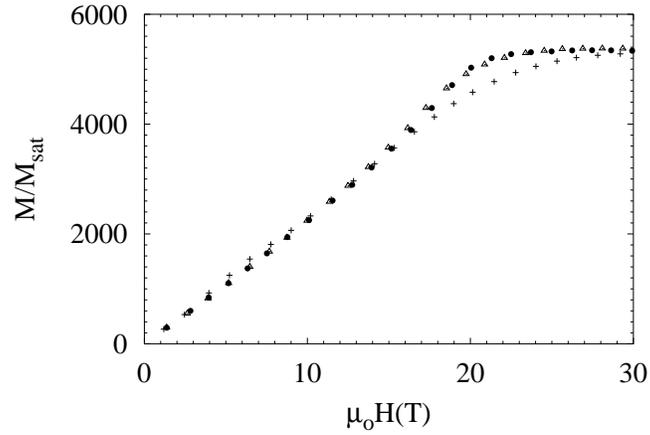}}
\caption{Molar magnetization vs H (T)  at three different temperatures for 
a powder sample of (5MAP)$_2$CuBr$_4$. T = 1.3 K:  ({\large$\bullet$}), 
T = 2.4 K: ($\triangle$), T = 4.4 K: (+).}
\label{map_mvh}
\end{figure}

\begin{figure}[h]
{\includegraphics[width=2.5in,angle=-90]{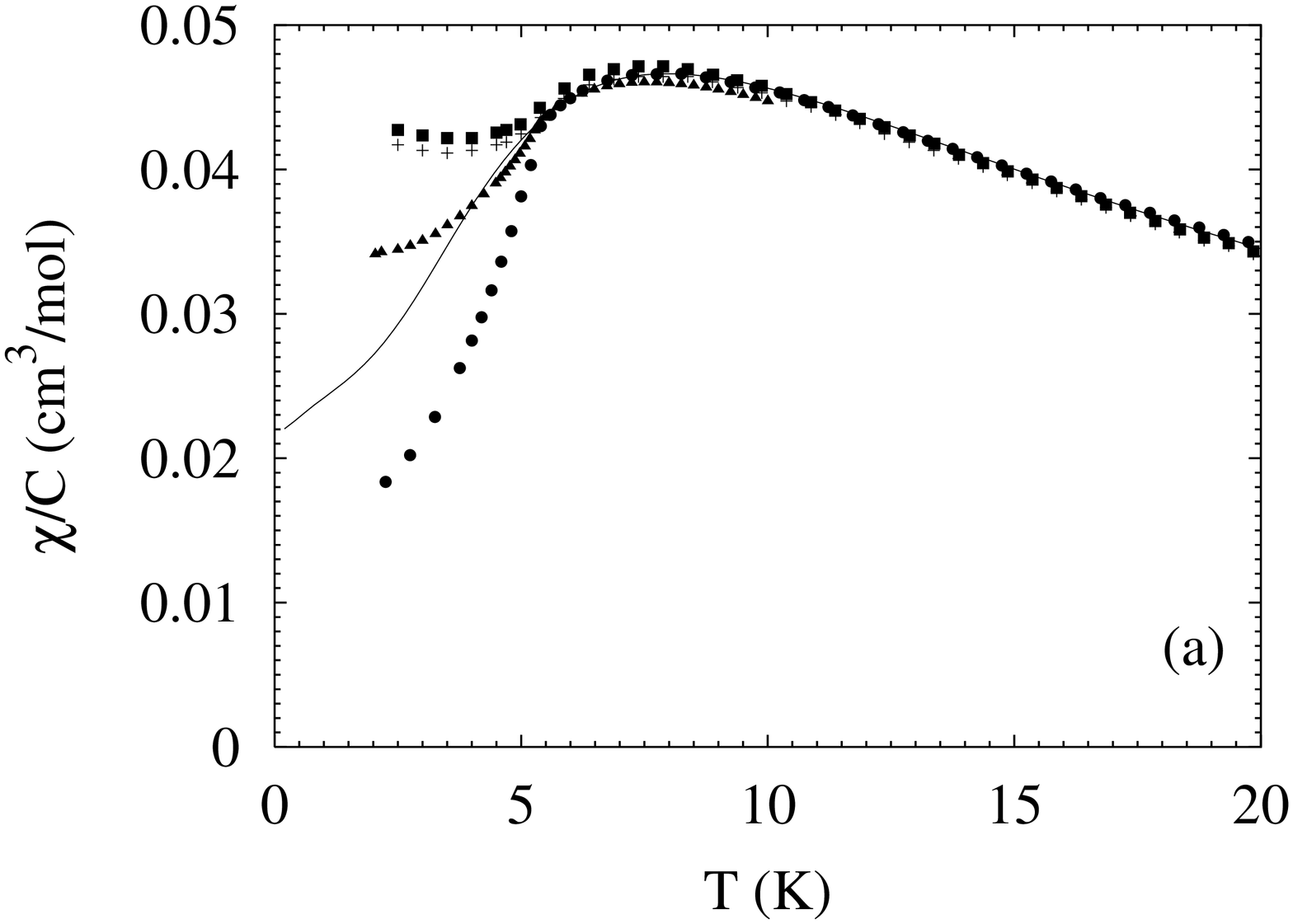}}
{\includegraphics[width=2.5in,angle=-90]{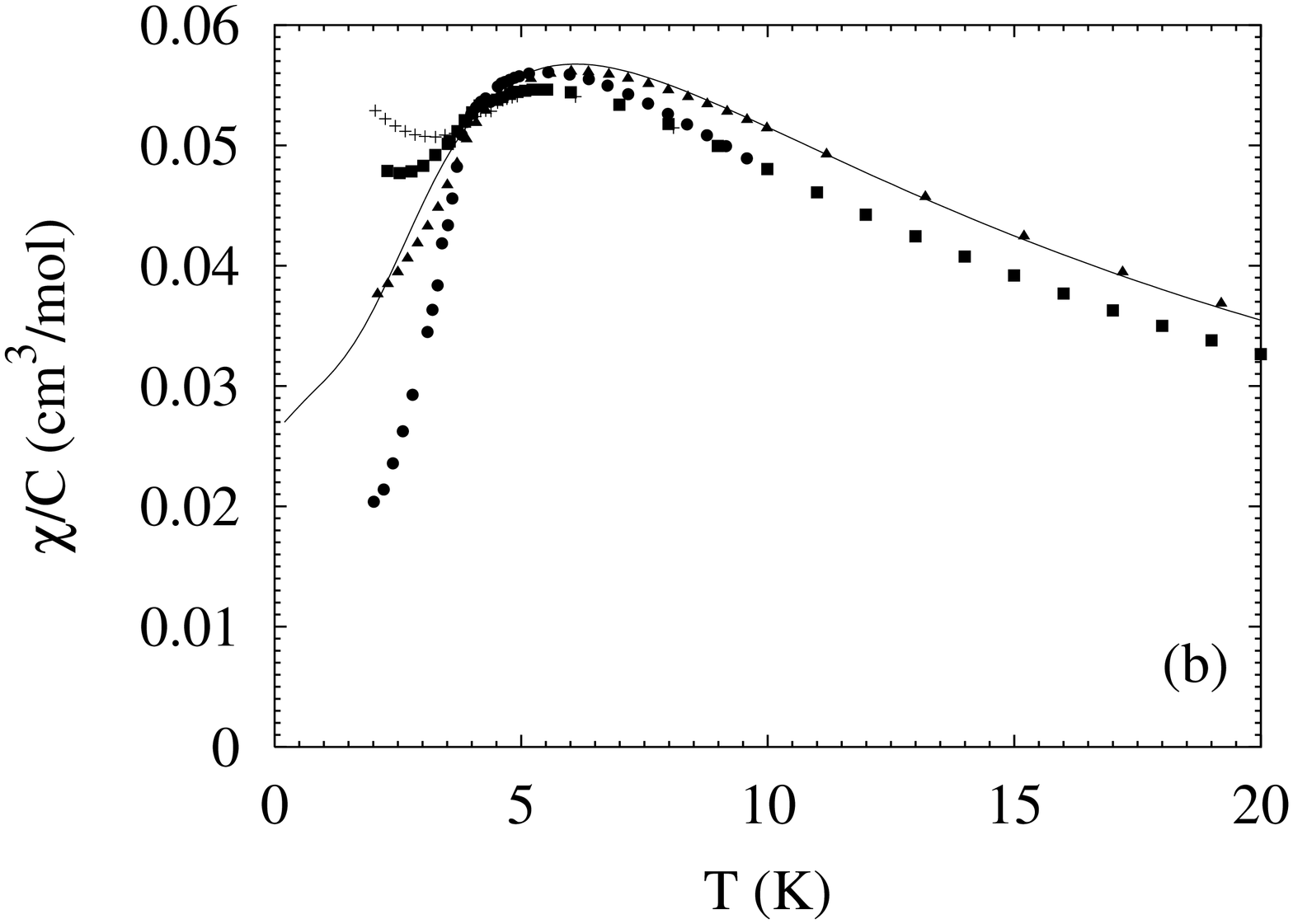}}
\caption{Single crystal $\chi$/C versus temperature at H = 0.2 T. 
(a):  (5CAP)$_2$CuBr$_4$, (b): (5MAP)$_2$CuBr$_4$. $\chi_{\perp}$: $\blacksquare$ H applied 
$\|$ to {\it b} and $\blacktriangle$ H applied $\|$ to {\it c}, 
$\chi_\|$: {\large$\bullet$} H applied $\|$ to {\it a}, $\chi_{powder}$: {\bf +} H = 0.1 T.
Solid line represents 2D QHAF model.}
\label{cxtal}
\end{figure}

\begin{figure}[h]
{\includegraphics[width=2.5in,angle=-90]{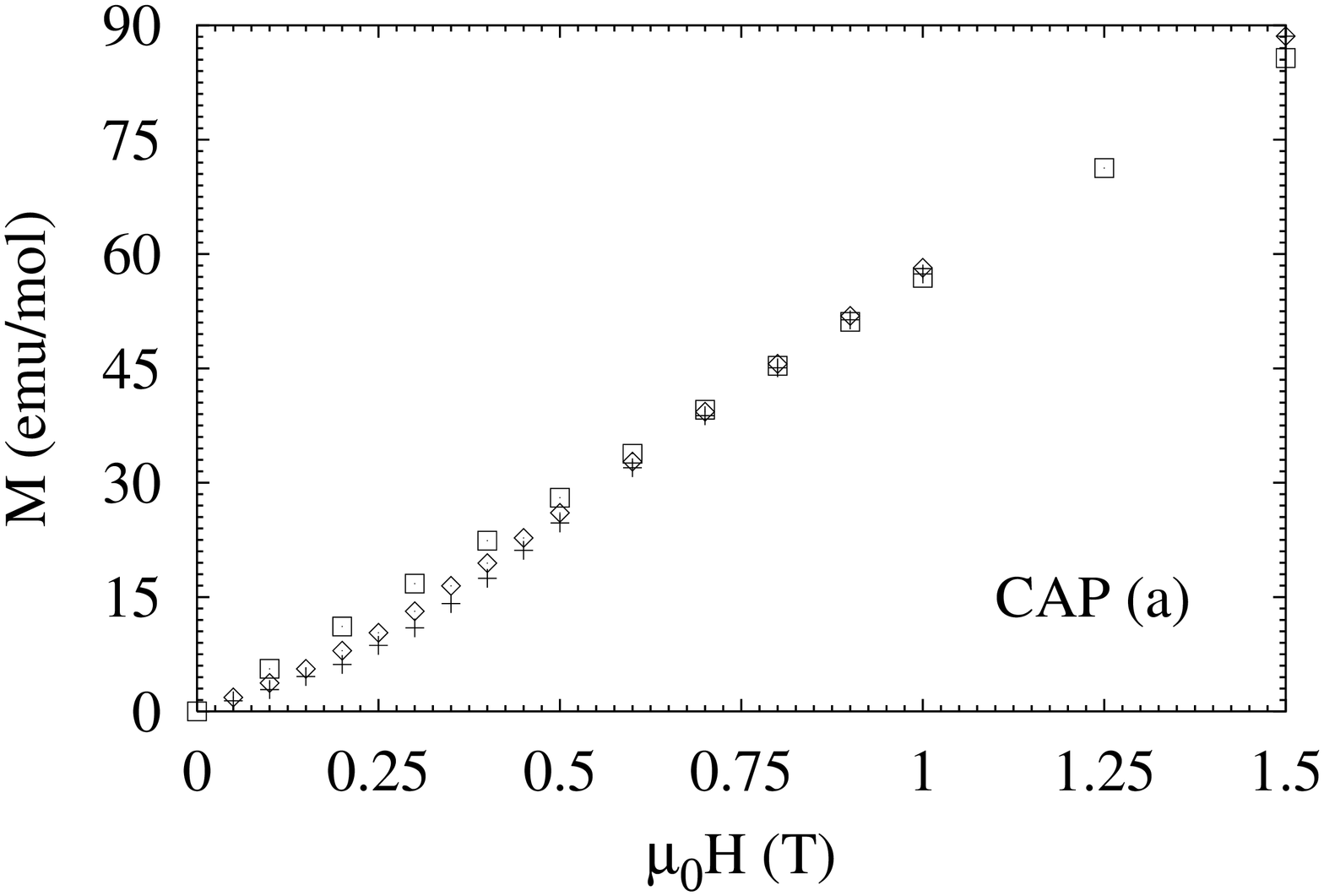}}
{\includegraphics[width=2.5in,angle=-90]{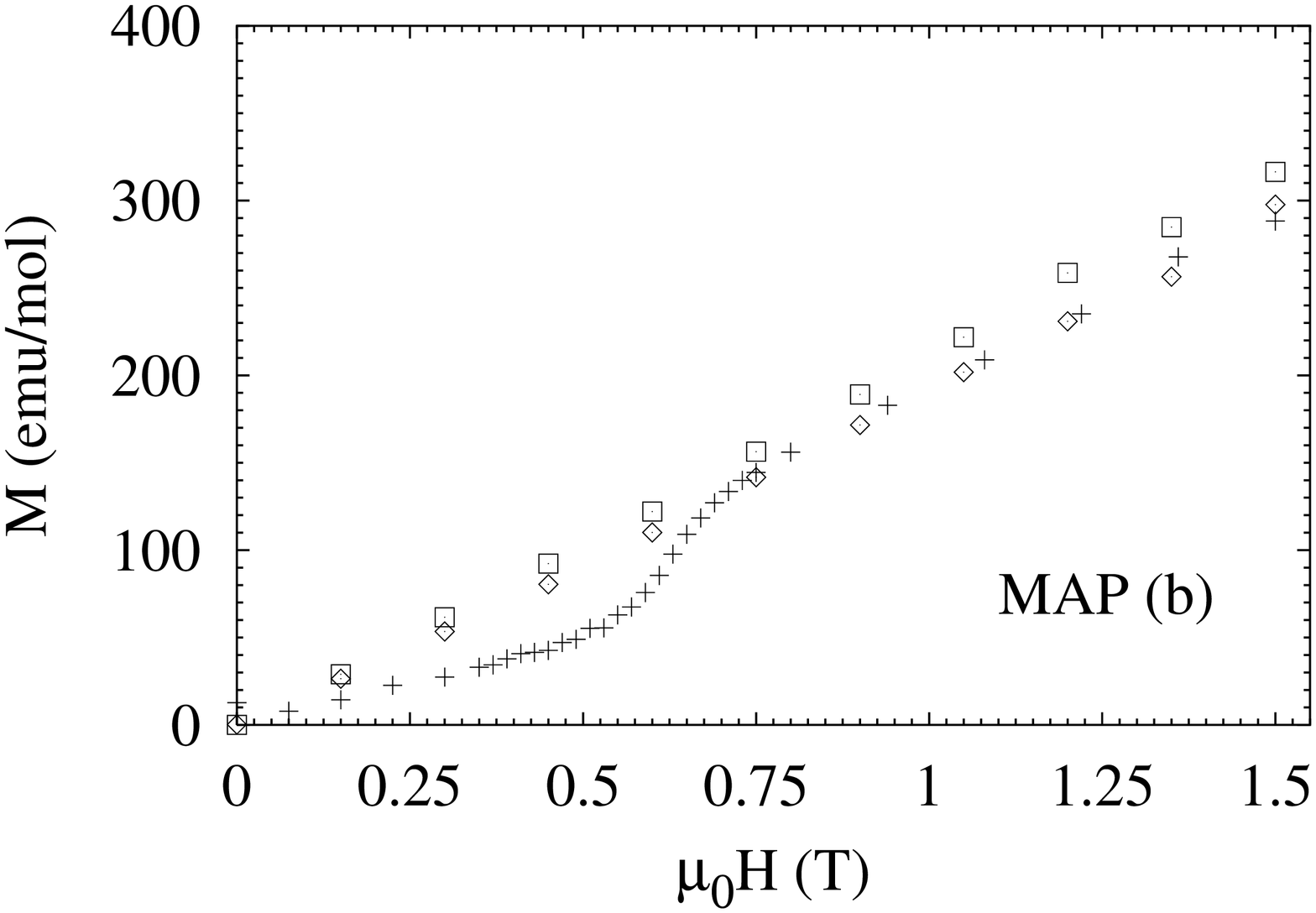}}
\caption{Single crystal molar magnetization at T = 2.1 K.  (a) (5CAP)$_2$CuBr$_4$ 
{\bf +} H applied $\|$ {\it a},  $\diamond$ H applied $\|$ {\it b}, $\square$ H applied $\|$
{\it c}. (b) (5MAP)$_2$CuBr$_4$ {\bf +} H applied $\|$ {\it a},  $\diamond$ H applied 
$\|$ {\it b}, $\square$ H applied $\|$
{\it c}.}
\label{mxtal}
\end{figure}

\begin{figure}[h]
{\includegraphics[width=2.5in,angle=-90]{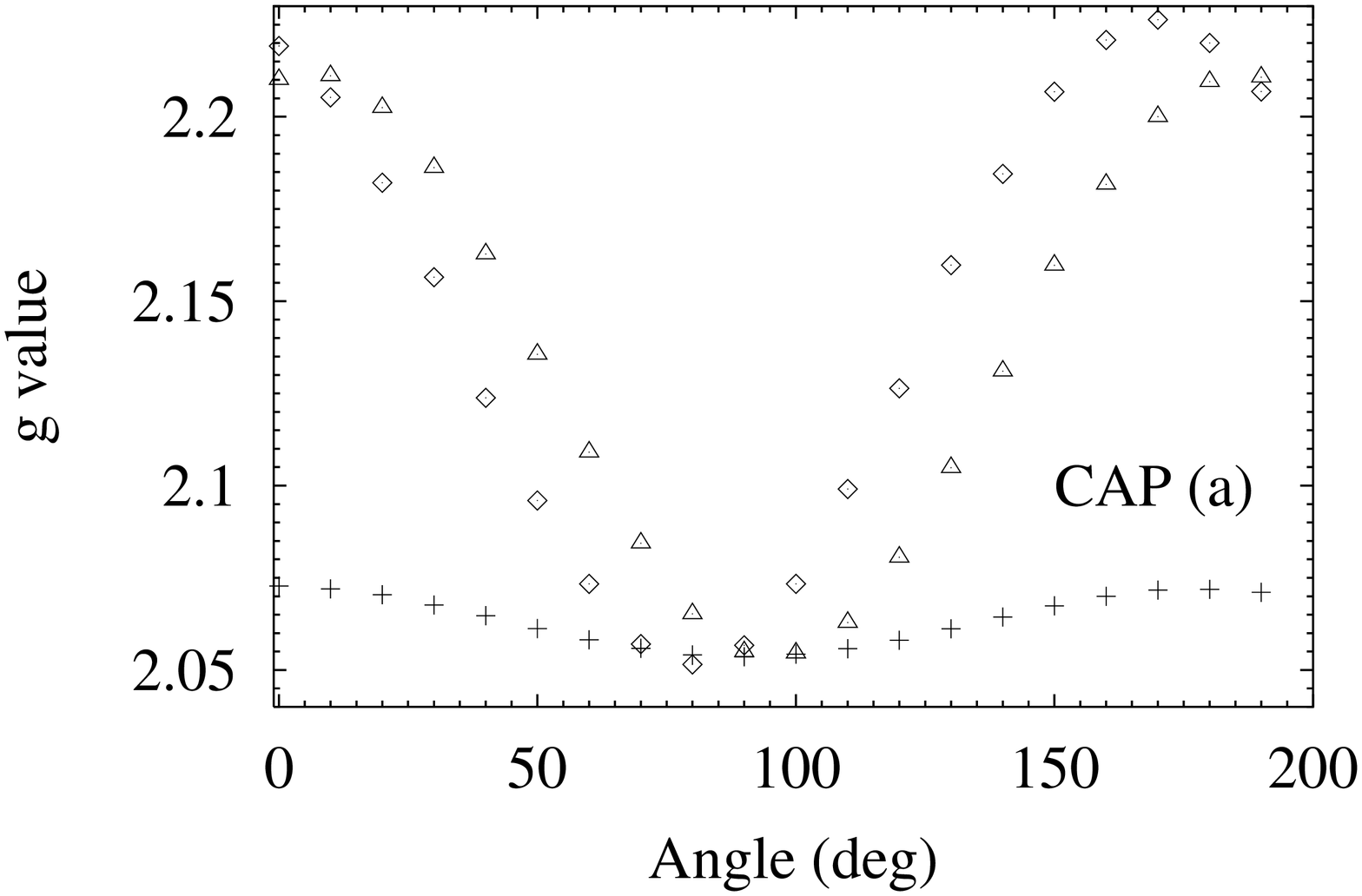}}
{\includegraphics[width=2.5in,angle=-90]{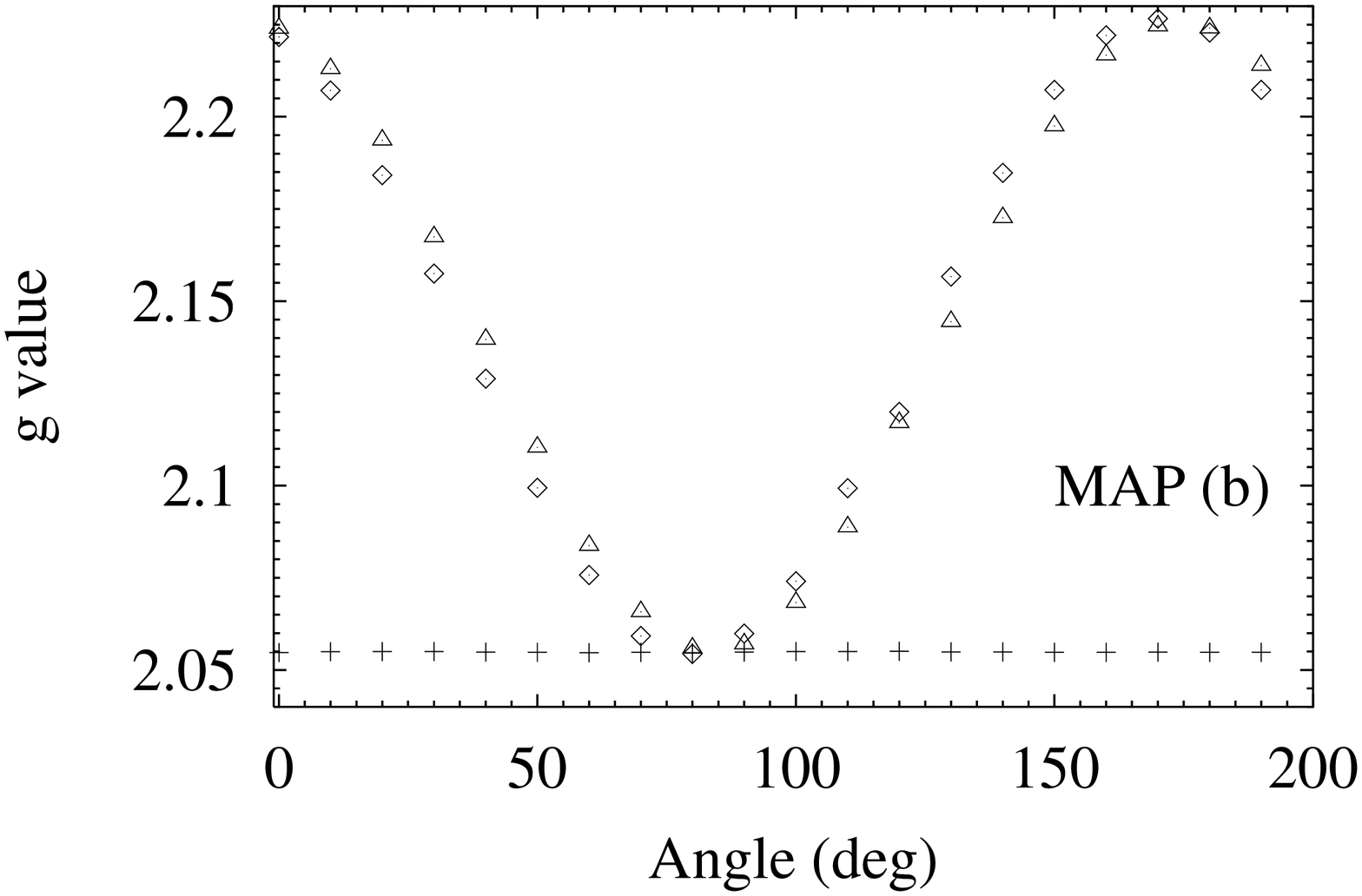}}
\caption{Single crystal, room temperature, X-band EPR for (5MAP)$_2$CuBr$_4$ 
in (a) and (5CAP)$_2$CuBr$_4$ in (b).  In both cases the {\bf +}
represent data for rotations in ab plane about the c* axis while $\triangle$ 
is for rotations in the ac* plane and $\diamond$ is for rotations in the bc*
plane.}
\label{eprsc}
\end{figure}

\begin{figure}[h]
{\includegraphics{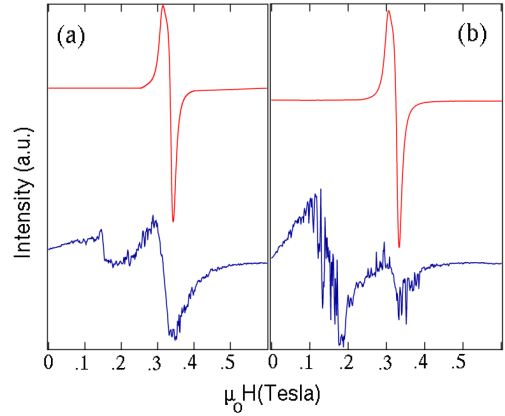}}
\caption{Powder X-band EPR for (5MAP)$_2$CuBr$_4$ (a) and
(5CAP)$_2$CuBr$_4$ (b).  The top spectra are at room temperature
and the bottom spectra were collected on an Oxford ESR910 He cryostat at 
3.2 K.}
\label{eprlt}
\end{figure}

{The low value of the exchange strength found for these 2D QHAF compounds 
allows the antiferromagnetic interactions to be overcome by experimentally 
accessible magnetic fields.  The required saturation fields can be estimated at 
T = 0 K by a mean field calculation.  Assuming the critical field depends only on 
the exchange strength, the equation for an S = 1/2 system is given 
by\cite{Bonner64}

\begin{equation}
H_{sat}={{zJ}\over{g\mu_{\beta}}},
\label{hsat}
\end{equation}
where $z$ is the number of nearest neighbors.  Therefore at T = 0 K, 
the predicted values of the saturation field for (5MAP)$_2$CuBr$_4$ and 
(5CAP)$_2$CuBr$_4$ are 18.8 T and 24.1 T, based upon the values for $J$ 
obtained from the fits to the susceptibility data (Figure \ref{mvh}(b)).  The 
fact that (5MAP)$_2$CuBr$_4$ is observed to saturate at fields lower than 
(5CAP)$_2$CuBr$_4$ is consistent with the smaller exchange strength 
(Figure \ref{mvh}(a)).  These predicted saturation values are in good agreement 
with the experimental data shown in Figures \ref{mvh} and \ref{map_mvh}.}

\par{The values of M$_{sat}$ for the high field magnetization are in good agreement
with the g-values known from powder EPR.  Given that

\begin{equation}
M_{sat} = g \mu_\beta N S
\end{equation}
and using either the powder EPR g-values (2.11 and 2.12) or the 2D QHAF model fit g-vaules
(2.10 and 2.07) the difference between the measured M$_{sat}$ (5980 and 5880 emu/mol)
and the calculated M$_{sat}$ (5900 emu/mol and 5905 emu/mol) is on the order of 1 \% 
(1.37 and 0.45 \%).}

\par{A noteworthy feature in the field dependent magnetization data contained 
in Figures \ref{mvh} and \ref{map_mvh} is the upward curvature of 
the low temperature magnetization data. This  behavior is qualitatively 
similar to theoretical\cite{Griffiths64,Bonner64} and experimental
\cite{Miura95} realizations of the T = 0 K magnetization curve of the one 
dimensional S = 1/2 Heisenberg antiferromagnet, indicated by the dashed curve 
in Figure \ref{mvh}.  The theory for the 1D QHAF also predicts that for relative 
temperatures kT/J $<$ 0.5, the magnetization has positive curvature prior to 
saturation. However, it is clear that the curvature in magnetization curve of 
the 1D model is more extreme than found experimentally for (5CAP)$_2$CuBr$_4$ 
and (5MAP)$_2$CuBr$_4$.}

\par{Calculations for the magnetization curve of the 2D QHAF have recently 
appeared\cite{Yang97,Zhitomirsky98,Troyer}.  These calculations have been 
based on a diagonalization of finite lattices at T = 0 K\cite{Yang97}, a  
T = 0 K spin-wave expansion with second order corrections \cite{Zhitomirsky98},
and Quantum Monte Carlo studies on large systems at both zero and finite 
temperatures\cite{Troyer}. All three sets of predictions are in good agreement with each other
at T = 0 K, but the two more recent papers \cite{Zhitomirsky98,Troyer} contain more precise 
predictions. The T = 0 K spin-wave expansion\cite{Zhitomirsky98} is represented as the 
dotted line, the Quantum Monte Carlo calculations\cite{Troyer} as the long dashed line, and
the 1D QHAF prediction\cite{Bonner64} as a short dashed line in Figure \ref{mvh}(b), with 
the less precise data of Yang et al\cite{Yang97} not shown. The field axis for the predictions was 
scaled by saturation fields based upon the experimental exchange strengths as determined by
susceptibilities (Eqn. \ref{hsat}). The data are in much better quantitative agreement with 
the 2D predictions than the 1D. We emphasize that the agreement between theory and 
experiment in Figure \ref{mvh}(b) is $not$ due to a fit, since no parameters were 
allowed to vary.}

\par{The agreement between the experimental data and theory is poorer
for low fields, but becomes better as the saturation field is approached. We 
attribute the low field discrepancies to a combination of the 3D ordering 
and finite temperature effects. In the 3D ordered state, the low temperature 
ratio of M/H in fields above the spin-flop field are nearly double the values 
predicted for the ideal 2D QHAF susceptibility. The initial slope of the theoretical 
magnetization curve is derived from the isolated layer model, and for this reason the 
slope is consistently lower than the experimental magnetization data. As the field increases 
toward the saturation value, the weak 3D interactions become increasingly irrelevant and the 
data fall onto the theoretical curve.}

\par{At the highest fields, the effects of finite temperatures are observed. 
The T = 0 K magnetization curve is predicted to have a weak logarithmic 
divergence at H$_{sat}$ due to the quenching of quantum fluctuations
\cite{Zhitomirsky98}, but this feature is not observed because the high field
magnetization curve is rounded due to the presence of thermal excitations. The 
magnetization curve M(H,T) has recently been calculated for several finite temperatures
by Quantum Monte-Carlo techniques\cite{Troyer}; these results are presented in Figure 
\ref{mvh}(b) as the long dashed lines corresponding to the relative temperature T/J = 
0.2. The scaled temperatures for the (5MAP)$_2$CuBr$_4$ and (5CAP)$_2$CuBr$_4$ compounds 
are T/J = 0.19 and 0.15, respectively.}

\par{Similar magnetization behavior has been observed experimentally 
\cite{Hammar97} for the analogous chloride complexes, (5CAP)$_2$CuCl$_4$ and 
(5MAP)$_2$CuCl$_4$. The isostructural chloride compounds have significantly weaker 
exchange interactions because of the smaller Cl radius resulting in a diminished
halide-halide overlap.  The low temperature magnetization curves for the 
chloride analogs are similar in shape to those observed for the bromides, but 
the saturation fields are considerably smaller (3.8 T and 2.4 T, respectively), which is 
consistent with the smaller exchange strengths (1.14 K and 0.76 K, respectively).}

\par{Low field single crystal magnetization data for (5CAP)$_2$CuBr$_4$ and 
(5MAP)$_2$CuBr$_4$ are presented in Figure \ref{mxtal}. The (5MAP)$_2$CuBr$_4$ magnetization
data exhibits a definite change of slope at H = 0.63 T, Figure \ref{mxtal}(a), when the 
applied field is parallel to the nominal easy axis.  The same behavior is seen in the 
(5CAP)$_2$CuBr$_4$ data, Figure \ref{mxtal}(b), in which the change of slope appears at 
H = 0.38 T when the field is applied parallel to the nominal easy axis.  When the applied
field is perpendicular to the easy axis, for both the (5CAP)$_2$CuBr$_4$ and 
(5MAP)$_2$CuBr$_4$, Figure \ref{mxtal}, the magnetization data are linear.  These 
data sets were all collected at T = 2.1 K, well below T$_N$ for both
compounds.} 

\par{The change in slope exhibited by these magnetization data is evidence of a 
spin-flop\cite{dejongh-miedema} transition due to a weak internal anisotropy field, 
H$_{aniso}$.  The strength of the anisotropy field relative to the primary exchange
field can be determined by mean-field arguments\cite{dejongh-miedema} given the
spin-flop field, H$_{sf}$, and the saturation magnetization field, H$_{sat}$ 

\begin{equation}
H_{aniso} \approx {{H_{sf}^2}\over{2 H_{sat}}}
\label{aniso}
\end{equation}
Based on Eqn. \ref{aniso}, the anisotropy fields for (5CAP)$_2$CuBr$_4$ and 
(5MAP)$_2$CuBr$_4$ are 0.0030 T and 0.0075 T respectively.  Arguments given
below place the value of the 3D exchange at J' $\approx$ 0.08J for both compounds.
When J' is compared to the anisotropy field, where H$_{aniso}$ $\approx$ 0.0004H$_{ex}$, 
clearly J' is a stronger interaction.  This implies that J' is the dominant interaction that
drives the system from a 2D QHAF to a 3D QHAF with a weak Ising anisotropy.}

\subsection{Three-Dimensional Interactions}
	
\par{Ultimately, a transition to long-range order will occur for all 
two-dimensional Heisenberg antiferromagnets at low enough temperatures.  This 
transition can be brought about either by the presence of small amounts of 
anisotropy (XY or Ising) or by magnetic interactions J$^{\prime}$ between the 
magnetic planes\cite{Navarro90}. A useful parameter for characterizing 
low-dimensional magnetic systems is the ratio of the temperature at which
long range order occurs to the interaction energy, T$_N$/$J$.  The properties 
of a number of 2D QHAF are compared in Table \ref{2ds} where it is noted that 
even the best isolated\cite{Greven95} 2D QHAFs become 3D ordered when the 
T$_N$/$J$ ratio has dropped to a value of $\geq$ 0.17. Values between 0.2 and 
0.4 have previously been found for other compounds which exhibit properties of 
2D QHAFs.}

\par{The transition to long-range order in (5CAP)$_{2}$CuBr$_{4}$ has been 
previously determined by specific heat studies\cite{Sorai96}. The specific 
heat of this compound shows a sharp maximum at 5.08 K which has been 
attributed to a magnetic ordering transition.  Analysis of the specific heat 
data above T$_{c}$ show  its  behavior to be that of a 2D QHAF with an 
exchange constant of 8.6 K, in excellent
agreement with the value obtained from the powder magnetic susceptibility
studies. The ratio of T$_c$/$J$ for 
(5CAP)$_{2}$CuBr$_{4}$ is therefore 0.60 (Table \ref{2ds}). While this value 
is higher than those found for the other 2D QHAF reported in Table \ref{2ds}, 
we note that it is not high enough to prevent the observation of the 
characteristic rounded maximum in the magnetic susceptibility or of the 
characteristic upward curvature in the magnetization curve.}
 
\par{The powder magnetic susceptibility data of (5CAP)$_{2}$CuBr$_{4}$ 
also show evidence of 3D ordering.  The data are well described by the 2D QHAF 
model of the susceptibility at temperatures above T$_{N}$ but deviate from the model curve 
at lower temperatures, Figure \ref{cap_chi}, clearly defining the Ne\'{e}l transition.
The powder susceptibility data collected in fields below the spin-flop field, H = 0.1 T, 
show only a small systematic deviation from the 2D model, first dropping below the model 
at T$_N$ then curving up to cross it at lower temperatures as the data approaches the 
powder average.  The powder data collected in a field of 1.0 T, above the spin-flop field,
show a more pronounced deviation from the 2D model at T$_N$.  Assuming the same relation
between the onset of 3D order and deviation of powder data from the 2D QHAF model, the 
critical temperature for (5MAP)$_{2}$CuBr$_{4}$ can be estimated from Figure \ref{map_chi}
to be 3.8(2) K which would correspond to a critical ratio of T$_{N}$/$J$ = 0.58.}

\par{The specific heat measurements\cite{Sorai96} of (5CAP)$_{2}$CuBr$_{4}$ are more sensitive 
to the crossover from 2D QHAF behavior to the 3D ordered state than the susceptibility data. The
magnetic specific heat is clearly higher than the prediction for the 2D model with J/k = 8.5 K for 
temperatures as high as 6.5 K (Fig.4 in Ref. 11), a full 30\% above T$_{N}$.}

\par{A more sensitive indicator of the degree of isolation of the 2D layers 
is the value of the correlation length $\xi$ at the critical ratio T$_N$/J. 
According to the theory of Chakravarty, Halprin, and Nelson\cite{Chakravarty88}, 
the correlation length diverges exponentially at low temperatures, with only a 
weak temperature dependence in the prefactors\cite{Hasenfratz93}. The full 
expression for the correlation length is} 

\begin{equation}
{\xi\over{a}}={e\over{8}}{c/a\over{2\pi\rho_{s}}}exp({2\pi\rho_{s}\over{T}})
(1-0.5{T\over{2\pi\rho_{s}}}+O({T\over{2\pi\rho_{s}}})^{2})
\label{cor}
\end{equation}

\noindent{where $c$=1.657Ja and $\rho_s$=0.1800J\cite{beard} are the renormalized spin wave 
velocity and spin-stiffness constants, respectively, and the correlation length is 
expressed in units of the lattice constant $a$.  For 5CAP and 5MAP  at their 
critical ratio T$_N$/J = 0.60, Equation \ref{cor} predicts the correlation 
length to be  $\xi$/a = 2.2. In comparison, the correlation length ratios 
($\xi$/a ) at T$_N$ for Sr$_{2}$CuO$_{2}$Cl$ _{2}$ and deuterated copper formate 
tetrahydrate have been determined by neutron scattering experiments to 
be close to 220 (Greven\cite{Greven95}) and 55 (Ronnow\cite{Ronnow99}), respectively.}
	
\par{The large differences between the correlations lengths of the four 
compounds at the critical temperatures is a reflection of large differences 
in the relative ratios of interplanar to intraplanar exchange, J$^{\prime}$/J. 
According to the mean-field theory of magnetic ordering in low-dimensional 
magnets\cite {Villain77}, long range order will set in when the thermal energy 
is comparable to the interaction energy between blocks of correlated spins of 
the z neighboring layers

\begin{equation}
kT_{N}\approx zJ^{\prime}S(S+1)({\xi(T_{N})\over{a}})^{2}
\end{equation}
Using the value of 2.2 for $\xi$/a at T$_N$ yields a value for 
J$^{\prime}$/k  $\approx$ 0.14T$_N$ = 0.14 (0.60 J/K) = 0.08 J/K = 0.72 K
A similar value, J'/k = 1.0 K, was obtained by analysis of the spin-wave contribution to the 
magnetic specific heat in the ordered state\cite{Sorai00}. We view the two estimates to 
be equivalent, considering the approximations used in the two analyses.  The same ratio is 
found for 5MAP.  In contrast, the same calculation for Sr$_{2}$CuO$_{2}$Cl$ _{2}$ yields a 
J'/J ratio of 1.2x10$^{-6}$}

\par{Why is the J$^{\prime}$/J ratio so large in the 5CAP/5MAP family of 
2D QHAFs? One important reason lies in the relationship between copper sites 
in two adjacent layers. For the well isolated systems 
(La$_2$CuO$_4$, Sr$_{2}$CuO$_{2}$Cl$ _{2}$, copper formate), adjacent layers 
are staggered, with the copper sites in one layer are displaced by (1/2, 1/2) 
with respect to those in adjacent layers, placing the metal ions equidistant 
from four equivalent metal sites in the layer above. Not only does this 
displacement increase the interlayer Cu$\cdots$Cu distance, it also provides a 
net cancellation of the four antiferromagnetic interactions from the adjacent 
layer. To first order J$^{\prime}$ vanishes in staggered systems, and 3D order 
is actually brought about by the presence of weak anisotropy terms (Ising, XY, 
Dzyaloshinsky-Moria) in the Hamiltonian.}

\par{In contrast, the copper layers of the 5CAP/5MAP family lie directly above 
adjacent layers (Fig. 2).  Hence there is a much shorter interplanar 
Cu$\cdots$Cu distance and no cancellation of interactions terms takes place. 
The interlayer Cu$\cdots$Cu distance is only {7.88 \AA}, virtually the same as 
the intralayer Cu$\cdots$Cu distance of {7.86 \AA}. More important than the 
Cu-Cu distances are the Br$\cdots$Br contact distances between adjacent 
CuBr$_4^{2+}$ tetrahedra, since the exchange interaction occurs though the 
overlap of the bromide wave functions. Within the layer, the Br$\cdots$Br 
separation is {4.35 \AA}, with only one Br$\cdots$Br contact between any pair 
of copper sites in the layer, Fig. 2. Between the layers, the Br-Br separation 
is {4.83 \AA}, and there are two such contacts between each pair of 
interacting copper ions, Fig. 3.  The resulting J$^{\prime}$/J ratio of 0.08 
is therefore a consequence of two identical interlayer interactions, each 
equal to 0.04 J.}

\subsection{Comparisons to Other 2D QHAF}

\par{To date, the compounds studied which approximate the 2D QHAF are few. A 
brief summary of these systems is given in Table \ref{2ds}.  Along with 
La$_2$CuO$_4$,\cite{Endoh74} other compounds which have been studied include 
Sr$_{2}$CuO$_{2}$Cl$ _{2}$,\cite{Greven95} copper formate(CuF4H),
\cite{Clarke92b,Ronnow99} copper formate urea(CuFUH),\cite{Yamagata83} and 
copper fluoride dihydrate.\cite{Koyama87} Major studies of La$_2$CuO$_4$ 
include the neutron scattering experiments of Hayden 
\em{et al.}\em\cite{Hayden91}  Much attention has  been given to the 
complicated N\'{e}el transition near 300 K.\cite{Maclaughlin94}  This 
magnetic  transition is complicated by the structural  transition near 500 K 
from tetragonal to orthorhombic symmetry which  causes a canting of the local 
CuO$_{6}$ octahedra, therefore creating both a  small intraplanar XY and 
Dzyaloshinsky-Moria anisotropy. Knowledge of  these anisotropies is essential 
to  understanding the theory \cite{Chakravarty91} and experiment \cite{Imai93} 
of the relaxation processes.}

\par{There is an absence of canting in Sr$_2$CuO$_2$Cl$_2$ which makes 
it a better realization of a 2D QHAF. A determination of the temperature
dependence of the correlation length\cite{Greven95} found essentially
perfect agreement between the data and the renormalized classical theory
of Chakravarty \em{et al.}\em\cite{Chakravarty88}  However, no evidence was 
seen for the predicted crossover from the renormalized classical regime to
quantum critical behavior at higher temperatures.}

\par{The non-oxide versions of the 2D QHAFs are characterized by much smaller 
exchange. This is an important advantage since certain experiments (EPR 
linewidth, magnetic specific heat, magnetization) are difficult to conduct on 
the oxides. However, each of these model compounds has its own set of 
limitations. The formate-based compounds, CuF4H and CuFUH, have two copper 
sites significantly canted one to another. Although the only study of the EPR 
linewidth divergence has been done on CuF4H,\cite{Castner93} its conclusions 
are rendered suspect by the anisotropy contributions to the relaxation 
process. Copper fluoride dihydrate has no canting, but does have significant 
3D interactions, as judged by its relatively high T$_{N}$/$J$ ratio.
\cite{Abrahams62}}

\par{The scarcity of well characterized realizations of 2D QHAF demonstrates a 
clear need for new magnets with which to probe the behavior of this important 
class of quantum magnets.  Desirable characteristics of the new materials 
will include  moderate exchange strengths, high local symmetries, and well 
isolated magnetic layers. Our initial results presented here indicate that 
(5CAP)$_2$CuBr$_4$, and (5MAP)$_2$CuBr$_4$ are good candidates for studying 
the properties of 2D QHAFs.  These systems have magnetic interactions that 
are in a desirable range for a variety of experiments.  Although there is a 
transition to long range order at relatively high temperature, temperatures 
above and below T$_N$ may still show predominant 2D QHAF behavior as 
demonstrated by the agreement of our magnetization data with calculations of 
finite systems (see Figure \ref{mvh}(b)).}  

\par{From the perspective of molecular-based magnetism, what can be done to 
improve the 5CAP/5MAP family of 2D QHAFs? It would be be good to lower the 
J$^{\prime}$/J ratio so the interpretation of experimental data will be less 
affected by 3D crossover effects. Since the value of J$^{\prime}$ is determined 
by the interlayer Br$\cdots$Br contacts, increasing the interlayer separation 
should have a dramatic impact upon J$^{\prime}$, without significant influence 
on J. The interlayer spacing $1\over2$$c$ is primarily determined by the 
length of the organic cations,  from the 2-amino group to the substituent in 
the 5-position, Fig 3.  Replacing the chlorine ion in 5CAP with a larger ion 
(bromide, iodine, cyanide) may serve to push the layers further apart. 
Structural studies\cite{Turnbull99} with the 5-bromo substituent 
have shown the interlayer Br-Br contact distance has expanded to {4.99(6) \AA} 
from {4.83 \AA} in (5CAP)$_2$CuBr$_4$ , while the contact distance within the 
layers has only increased to {4.396 \AA}  from {4.35 \AA} (found in the 
(5CAP)$_2$CuBr$_4$). Increasing the size of the 5-substituent led to changes in the key Br-Br
contact distances which caused the expected changes in the magnetism
for the (5BAP)$_2$CuBr$_4$ compound\cite{5bap-art}.  Changes in the Br-Br contacts
resulted in J = 6.9(1) K, T$_N$ = 3.8(2) K causing a moderate reduction in the 
T$_N$/J ratio to 0.57.}
\par{Increasing the size of the 5-substituent of the pyridine ring still further,
by use of iodine, forces the complex to a completely new structure.  The compound 
(5IAP)$_2$CuBr$_4\bullet$2H$_2$O consists of ladders of close packed CuBr$_4^{2-}$ groups
with magnetic interactions J$_{rung}$ = 13 K and J$_{rail}$ = 1 K\cite{IAP}.
}

\par{Another family of 2D QHAF, with better 3D isolation than the CuBr$_4$ compounds, is 
also under investigation by our group.  These compounds consist of Cu$^{2+}$
ions mutually linked by neutral pyrazine molecules (pz, C$_4$H$_4$N$_2$) into 
antiferromagnetic layers\cite{Turnbull99,pz-bf4}:  Cu(pz)$_2$(ClO$_4$)$_2$, 
Cu(pz)$_2$(BF$_4$)$_2$, and $[$Cu(pz)$_2$(NO$_3$)](PF$_6$).  The exchange strengths, 
10.5 - 17 K, are stronger and the 3d isolation is better, T$_N$/J $\approx$ 0.25, than
the CuBr$_4$ compounds. The improvement in the critical ratio is due to the wide 
separation of the layers by the interleaved anions.  Full reports of synthesis, 
structures, and magnetic properties of these compounds are in preparation.}

\section{Acknowledgments}
\par{The authors would like to acknowledge Ward Robinson, University of 
Canterbury, for assistance with the X-ray data collection, Dan Reich, Johns 
Hopkins University, and Joseph Budnick, University of Connecticut-Storrs, for the use of 
their SQUID magnetometers, and M. Sorai, Microcalorimetry Center - Osaka University, for his 
work on the specific heat. We had useful discussions with E. Manousakis and 
Martin Graven as well as M. Troyer who has let us use his unpublished results on the
magnetization curve of the 2D QHAF.  A portion of this work was performed at the National High 
Magnetic Field Laboratory, which is supported by SF Cooperative Agreement No. 
DMR-9527035 and by the State of Florida. We received support from the NSF 
through grant DMR-9006470 and DMR-9803813.}

\bibliography{5xap}

\begin{thebibliography}{52}
\expandafter\ifx\csname natexlab\endcsname\relax\def\natexlab#1{#1}\fi
\expandafter\ifx\csname bibnamefont\endcsname\relax
  \def\bibnamefont#1{#1}\fi
\expandafter\ifx\csname bibfnamefont\endcsname\relax
  \def\bibfnamefont#1{#1}\fi
\expandafter\ifx\csname citenamefont\endcsname\relax
  \def\citenamefont#1{#1}\fi
\expandafter\ifx\csname url\endcsname\relax
  \def\url#1{\texttt{#1}}\fi
\expandafter\ifx\csname urlprefix\endcsname\relax\def\urlprefix{URL }\fi
\providecommand{\bibinfo}[2]{#2}
\providecommand{\eprint}[2][]{\url{#2}}

\bibitem[{\citenamefont{Birgeneau}(1990)}]{Birgeneau90}
\bibinfo{author}{\bibfnamefont{R.~J.} \bibnamefont{Birgeneau}},
  \bibinfo{journal}{Am. J. Phys} \textbf{\bibinfo{volume}{58}},
  \bibinfo{pages}{28} (\bibinfo{year}{1990}).

\bibitem[{\citenamefont{Sokol and Pines}(1993)}]{Sokol93}
\bibinfo{author}{\bibfnamefont{A.}~\bibnamefont{Sokol}} \bibnamefont{and}
  \bibinfo{author}{\bibfnamefont{D.}~\bibnamefont{Pines}},
  \bibinfo{journal}{Phys. Rev. Lett.} \textbf{\bibinfo{volume}{71}},
  \bibinfo{pages}{2813} (\bibinfo{year}{1993}).

\bibitem[{\citenamefont{Manousakis}(1991)}]{Manousakis91}
\bibinfo{author}{\bibfnamefont{E.}~\bibnamefont{Manousakis}},
  \bibinfo{journal}{Rev. of Mod. Phys.} \textbf{\bibinfo{volume}{63}},
  \bibinfo{pages}{1} (\bibinfo{year}{1991}).

\bibitem[{\citenamefont{Yang and M{\"u}tter}(1997)}]{Yang97}
\bibinfo{author}{\bibfnamefont{M.~S.} \bibnamefont{Yang}} \bibnamefont{and}
  \bibinfo{author}{\bibfnamefont{K.}~\bibnamefont{M{\"u}tter}},
  \bibinfo{journal}{Z. Phys. B} \textbf{\bibinfo{volume}{104}},
  \bibinfo{pages}{117} (\bibinfo{year}{1997}).

\bibitem[{\citenamefont{Zhitomirsky and Nikuni}(1998)}]{Zhitomirsky98}
\bibinfo{author}{\bibfnamefont{M.~E.} \bibnamefont{Zhitomirsky}}
  \bibnamefont{and} \bibinfo{author}{\bibfnamefont{T.}~\bibnamefont{Nikuni}},
  \bibinfo{journal}{Phys. Rev. B} \textbf{\bibinfo{volume}{57}},
  \bibinfo{pages}{5013} (\bibinfo{year}{1998}).

\bibitem[{\citenamefont{Zhitomirsky and Chernyshev}(1999)}]{Zhitomirsky99}
\bibinfo{author}{\bibfnamefont{M.~E.} \bibnamefont{Zhitomirsky}}
  \bibnamefont{and} \bibinfo{author}{\bibfnamefont{A.~L.}
  \bibnamefont{Chernyshev}}, \bibinfo{journal}{Phys. Rev. L}
  \textbf{\bibinfo{volume}{82}}, \bibinfo{pages}{4536} (\bibinfo{year}{1999}).

\bibitem[{\citenamefont{Kahn}(1993)}]{Kahn93}
\bibinfo{author}{\bibfnamefont{O.}~\bibnamefont{Kahn}},
  \emph{\bibinfo{title}{Molecular Magnetism}} (\bibinfo{publisher}{VCH
  Publishers, Inc.}, \bibinfo{address}{N.Y., N.Y.}, \bibinfo{year}{1993}).

\bibitem[{\citenamefont{Place and Willett}(1987)}]{Place87}
\bibinfo{author}{\bibfnamefont{H.}~\bibnamefont{Place}} \bibnamefont{and}
  \bibinfo{author}{\bibfnamefont{R.}~\bibnamefont{Willett}},
  \bibinfo{journal}{Acta Cryst. C} \textbf{\bibinfo{volume}{43}},
  \bibinfo{pages}{1050} (\bibinfo{year}{1987}).

\bibitem[{\citenamefont{Zhou et~al.}(1991)\citenamefont{Zhou, Drumheller,
  Rubenacker, Halvorson, and Willett}}]{Zhou91}
\bibinfo{author}{\bibfnamefont{P.}~\bibnamefont{Zhou}},
  \bibinfo{author}{\bibfnamefont{J.}~\bibnamefont{Drumheller}},
  \bibinfo{author}{\bibfnamefont{G.}~\bibnamefont{Rubenacker}},
  \bibinfo{author}{\bibfnamefont{K.}~\bibnamefont{Halvorson}},
  \bibnamefont{and} \bibinfo{author}{\bibfnamefont{R.}~\bibnamefont{Willett}},
  \bibinfo{journal}{J. Appl. Phys.} \textbf{\bibinfo{volume}{69}},
  \bibinfo{pages}{5804} (\bibinfo{year}{1991}).

\bibitem[{\citenamefont{Willett et~al.}(1988)\citenamefont{Willett, Place, and
  Middleton}}]{Willett88}
\bibinfo{author}{\bibfnamefont{R.}~\bibnamefont{Willett}},
  \bibinfo{author}{\bibfnamefont{H.}~\bibnamefont{Place}}, \bibnamefont{and}
  \bibinfo{author}{\bibfnamefont{M.}~\bibnamefont{Middleton}},
  \bibinfo{journal}{J. Am. Chem. Soc} \textbf{\bibinfo{volume}{110}},
  \bibinfo{pages}{8639} (\bibinfo{year}{1988}).

\bibitem[{\citenamefont{Matsumoto et~al.}(2001)\citenamefont{Matsumoto,
  Miyazaki, Albrecht, Landee, Turnbull, and Sorai}}]{Sorai96}
\bibinfo{author}{\bibfnamefont{T.}~\bibnamefont{Matsumoto}},
  \bibinfo{author}{\bibfnamefont{Y.}~\bibnamefont{Miyazaki}},
  \bibinfo{author}{\bibfnamefont{A.~A.} \bibnamefont{Albrecht}},
  \bibinfo{author}{\bibfnamefont{C.~P.} \bibnamefont{Landee}},
  \bibinfo{author}{\bibfnamefont{M.~M.} \bibnamefont{Turnbull}},
  \bibnamefont{and} \bibinfo{author}{\bibfnamefont{M.}~\bibnamefont{Sorai}},
  \bibinfo{journal}{J Phys Chem B} \textbf{\bibinfo{volume}{104}},
  \bibinfo{pages}{9993} (\bibinfo{year}{2001}).

\bibitem[{\citenamefont{Turnbull et~al.}(1999)\citenamefont{Turnbull, Albrecht,
  Jameson, , and Landee}}]{Turnbull99}
\bibinfo{author}{\bibfnamefont{M.~M.} \bibnamefont{Turnbull}},
  \bibinfo{author}{\bibfnamefont{A.~S.} \bibnamefont{Albrecht}},
  \bibinfo{author}{\bibfnamefont{G.~B.} \bibnamefont{Jameson}}, ,
  \bibnamefont{and} \bibinfo{author}{\bibfnamefont{C.~P.}
  \bibnamefont{Landee}}, \bibinfo{journal}{Mol. Cryst. Liq. Cryst.}
  \textbf{\bibinfo{volume}{335}}, \bibinfo{pages}{245} (\bibinfo{year}{1999}).

\bibitem[{\citenamefont{Griffiths}(1964)}]{Griffiths64}
\bibinfo{author}{\bibfnamefont{R.~B.} \bibnamefont{Griffiths}},
  \bibinfo{journal}{Phys. Rev.} \textbf{\bibinfo{volume}{133}},
  \bibinfo{pages}{A768,} (\bibinfo{year}{1964}).

\bibitem[{\citenamefont{Troyer}(2000)}]{troyer2}
\bibinfo{author}{\bibfnamefont{M.}~\bibnamefont{Troyer}}
  (\bibinfo{year}{2000}), \bibinfo{note}{private communication}.

\bibitem[{\citenamefont{Poole and Farach}(1972)}]{Poole72}
\bibinfo{author}{\bibfnamefont{C.}~\bibnamefont{Poole}} \bibnamefont{and}
  \bibinfo{author}{\bibfnamefont{H.}~\bibnamefont{Farach}},
  \emph{\bibinfo{title}{The Theory of Magnetic Resonance}}
  (\bibinfo{publisher}{John Wiely and Sons, Inc.}, \bibinfo{address}{N.Y.,
  N.Y.}, \bibinfo{year}{1972}).

\bibitem[{\citenamefont{Hammar et~al.}(1997)\citenamefont{Hammar, Dender,
  Reich, Albrecht, and Landee}}]{Hammar97}
\bibinfo{author}{\bibfnamefont{P.}~\bibnamefont{Hammar}},
  \bibinfo{author}{\bibfnamefont{D.}~\bibnamefont{Dender}},
  \bibinfo{author}{\bibfnamefont{D.}~\bibnamefont{Reich}},
  \bibinfo{author}{\bibfnamefont{A.}~\bibnamefont{Albrecht}}, \bibnamefont{and}
  \bibinfo{author}{\bibfnamefont{C.}~\bibnamefont{Landee}},
  \bibinfo{journal}{J. Appl. Phys.} \textbf{\bibinfo{volume}{81}},
  \bibinfo{pages}{4615} (\bibinfo{year}{1997}).

\bibitem[{\citenamefont{Rushbrooke et~al.}(1974)\citenamefont{Rushbrooke,
  Baker, and Wood}}]{Rushbrooke74}
\bibinfo{author}{\bibfnamefont{G.~S.} \bibnamefont{Rushbrooke}},
  \bibinfo{author}{\bibfnamefont{G.~A.} \bibnamefont{Baker}}, \bibnamefont{and}
  \bibinfo{author}{\bibfnamefont{P.~J.} \bibnamefont{Wood}},
  \emph{\bibinfo{title}{Phase Transitions and Critical Phenomena}}
  (\bibinfo{publisher}{Academic Press}, \bibinfo{year}{1974}).

\bibitem[{\citenamefont{Takahashi}(1989)}]{Takahashi89}
\bibinfo{author}{\bibfnamefont{M.}~\bibnamefont{Takahashi}},
  \bibinfo{journal}{Phys. Rev. B} \textbf{\bibinfo{volume}{40}},
  \bibinfo{pages}{2494} (\bibinfo{year}{1989}).

\bibitem[{\citenamefont{Singh and Gelfand}(1990)}]{Singh90}
\bibinfo{author}{\bibfnamefont{R.~R.~P.} \bibnamefont{Singh}} \bibnamefont{and}
  \bibinfo{author}{\bibfnamefont{M.~P.} \bibnamefont{Gelfand}},
  \bibinfo{journal}{Phys. Rev. B} \textbf{\bibinfo{volume}{42}},
  \bibinfo{pages}{996} (\bibinfo{year}{1990}).

\bibitem[{\citenamefont{Kim and Troyer}(1998)}]{Troyer}
\bibinfo{author}{\bibfnamefont{J.-K.} \bibnamefont{Kim}} \bibnamefont{and}
  \bibinfo{author}{\bibfnamefont{M.}~\bibnamefont{Troyer}},
  \bibinfo{journal}{Phys. Rev. Lett.} \textbf{\bibinfo{volume}{80}},
  \bibinfo{pages}{2705} (\bibinfo{year}{1998}).

\bibitem[{\citenamefont{Makivi\'c and Ding}(1991)}]{Makivic91}
\bibinfo{author}{\bibfnamefont{M.~S.} \bibnamefont{Makivi\'c}}
  \bibnamefont{and} \bibinfo{author}{\bibfnamefont{H.-Q.} \bibnamefont{Ding}},
  \bibinfo{journal}{Phys. Rev. B} \textbf{\bibinfo{volume}{43}},
  \bibinfo{pages}{3562} (\bibinfo{year}{1991}).

\bibitem[{\citenamefont{Eggert et~al.}(1994)\citenamefont{Eggert, Affleck, and
  Takahashi}}]{Eggert94}
\bibinfo{author}{\bibfnamefont{S.}~\bibnamefont{Eggert}},
  \bibinfo{author}{\bibfnamefont{I.}~\bibnamefont{Affleck}}, \bibnamefont{and}
  \bibinfo{author}{\bibfnamefont{M.}~\bibnamefont{Takahashi}},
  \bibinfo{journal}{Phys. Rev. Lett.} \textbf{\bibinfo{volume}{73}},
  \bibinfo{pages}{332} (\bibinfo{year}{1994}).

\bibitem[{\citenamefont{Willett}(1986)}]{willet_epr}
\bibinfo{author}{\bibfnamefont{R.}~\bibnamefont{Willett}},
  \bibinfo{journal}{Inorg. Chem.} \textbf{\bibinfo{volume}{25}},
  \bibinfo{pages}{1918} (\bibinfo{year}{1986}).

\bibitem[{\citenamefont{Bonner and Fisher}(1964)}]{Bonner64}
\bibinfo{author}{\bibfnamefont{J.}~\bibnamefont{Bonner}} \bibnamefont{and}
  \bibinfo{author}{\bibfnamefont{M.}~\bibnamefont{Fisher}},
  \bibinfo{journal}{Phys. Rev.} \textbf{\bibinfo{volume}{135}},
  \bibinfo{pages}{A640,} (\bibinfo{year}{1964}).

\bibitem[{Miu(1995)}]{Miura95}
\emph{\bibinfo{title}{Low Dimensional Systems of Semiconductors and Magnetic
  Materials in Very High Magnetic Fields up to 500 T}}
  (\bibinfo{publisher}{World Scientific}, \bibinfo{address}{New Jersey},
  \bibinfo{year}{1995}).

\bibitem[{\citenamefont{Dejongh et~al.}(1972)\citenamefont{Dejongh, Amstel, and
  Miedema}}]{dejongh-miedema}
\bibinfo{author}{\bibfnamefont{L.~J.} \bibnamefont{Dejongh}},
  \bibinfo{author}{\bibfnamefont{W.~D.~V.} \bibnamefont{Amstel}},
  \bibnamefont{and} \bibinfo{author}{\bibfnamefont{A.~R.}
  \bibnamefont{Miedema}}, \bibinfo{journal}{Physica B}
  \textbf{\bibinfo{volume}{58}}, \bibinfo{pages}{277} (\bibinfo{year}{1972}).

\bibitem[{\citenamefont{Navarro}(1990)}]{Navarro90}
\bibinfo{author}{\bibfnamefont{R.}~\bibnamefont{Navarro}}, in
  \emph{\bibinfo{booktitle}{Magnetic Properties of Layered Transition Metal
  Compounds}}, edited by \bibinfo{editor}{\bibfnamefont{L.~D.}
  \bibnamefont{Jongh}} (\bibinfo{publisher}{Kluwer Academic Press},
  \bibinfo{address}{Dordecht}, \bibinfo{year}{1990}), p. \bibinfo{pages}{105}.

\bibitem[{\citenamefont{Greven et~al.}(1995)\citenamefont{Greven, Birgeneau,
  Endoh, Kastner, Matsuda, and Shirane}}]{Greven95}
\bibinfo{author}{\bibfnamefont{M.}~\bibnamefont{Greven}},
  \bibinfo{author}{\bibfnamefont{R.}~\bibnamefont{Birgeneau}},
  \bibinfo{author}{\bibfnamefont{Y.}~\bibnamefont{Endoh}},
  \bibinfo{author}{\bibfnamefont{M.}~\bibnamefont{Kastner}},
  \bibinfo{author}{\bibfnamefont{M.}~\bibnamefont{Matsuda}}, \bibnamefont{and}
  \bibinfo{author}{\bibfnamefont{G.}~\bibnamefont{Shirane}},
  \bibinfo{journal}{Z. Phys. B} \textbf{\bibinfo{volume}{96}},
  \bibinfo{pages}{465} (\bibinfo{year}{1995}).

\bibitem[{\citenamefont{Chakravarty et~al.}(1988)\citenamefont{Chakravarty,
  Halperin, and Nelson}}]{Chakravarty88}
\bibinfo{author}{\bibfnamefont{S.}~\bibnamefont{Chakravarty}},
  \bibinfo{author}{\bibfnamefont{B.~I.} \bibnamefont{Halperin}},
  \bibnamefont{and} \bibinfo{author}{\bibfnamefont{D.~R.}
  \bibnamefont{Nelson}}, \bibinfo{journal}{Phys. Rev. Lett.}
  \textbf{\bibinfo{volume}{60}}, \bibinfo{pages}{1057} (\bibinfo{year}{1988}).

\bibitem[{\citenamefont{Hasenfratz and Niedermayer}(1993)}]{Hasenfratz93}
\bibinfo{author}{\bibfnamefont{P.}~\bibnamefont{Hasenfratz}} \bibnamefont{and}
  \bibinfo{author}{\bibfnamefont{F.}~\bibnamefont{Niedermayer}},
  \bibinfo{journal}{Z. Phys B} \textbf{\bibinfo{volume}{92}},
  \bibinfo{pages}{91} (\bibinfo{year}{1993}).

\bibitem[{\citenamefont{Beard et~al.}(1998)\citenamefont{Beard, Birgeneau,
  Greven, and Wiese}}]{beard}
\bibinfo{author}{\bibfnamefont{B.~B.} \bibnamefont{Beard}},
  \bibinfo{author}{\bibfnamefont{R.~J.} \bibnamefont{Birgeneau}},
  \bibinfo{author}{\bibfnamefont{M.}~\bibnamefont{Greven}}, \bibnamefont{and}
  \bibinfo{author}{\bibfnamefont{U.-J.} \bibnamefont{Wiese}},
  \bibinfo{journal}{Phys. Rev. Lett.} \textbf{\bibinfo{volume}{80}},
  \bibinfo{pages}{1742} (\bibinfo{year}{1998}).

\bibitem[{\citenamefont{Ronnow et~al.}(1999)\citenamefont{Ronnow, McMorrow, and
  Harrison}}]{Ronnow99}
\bibinfo{author}{\bibfnamefont{H.~M.} \bibnamefont{Ronnow}},
  \bibinfo{author}{\bibfnamefont{D.~F.} \bibnamefont{McMorrow}},
  \bibnamefont{and} \bibinfo{author}{\bibfnamefont{A.}~\bibnamefont{Harrison}},
  \bibinfo{journal}{Phys. Rev. Lett} \textbf{\bibinfo{volume}{82}},
  \bibinfo{pages}{3152} (\bibinfo{year}{1999}).

\bibitem[{\citenamefont{Villain and Loveluck}(1977)}]{Villain77}
\bibinfo{author}{\bibfnamefont{J.}~\bibnamefont{Villain}} \bibnamefont{and}
  \bibinfo{author}{\bibfnamefont{J.~M.} \bibnamefont{Loveluck}},
  \bibinfo{journal}{J. de Physique Lett} \textbf{\bibinfo{volume}{88}},
  \bibinfo{pages}{L77} (\bibinfo{year}{1977}).

\bibitem[{\citenamefont{Matsumoto et~al.}(2000)\citenamefont{Matsumoto,
  Miyazaki, Albrecht, Landee, Turnbull, and Sorai}}]{Sorai00}
\bibinfo{author}{\bibfnamefont{T.}~\bibnamefont{Matsumoto}},
  \bibinfo{author}{\bibfnamefont{Y.}~\bibnamefont{Miyazaki}},
  \bibinfo{author}{\bibfnamefont{A.~S.} \bibnamefont{Albrecht}},
  \bibinfo{author}{\bibfnamefont{C.~P.} \bibnamefont{Landee}},
  \bibinfo{author}{\bibfnamefont{M.~M.} \bibnamefont{Turnbull}},
  \bibnamefont{and} \bibinfo{author}{\bibfnamefont{M.}~\bibnamefont{Sorai}},
  \bibinfo{journal}{J. Phys. Chem. B} \textbf{\bibinfo{volume}{104}},
  \bibinfo{pages}{9993} (\bibinfo{year}{2000}).

\bibitem[{\citenamefont{Endoh et~al.}(1974)\citenamefont{Endoh, Shirane,
  Birgeneau, Richards, and Holt}}]{Endoh74}
\bibinfo{author}{\bibfnamefont{Y.}~\bibnamefont{Endoh}},
  \bibinfo{author}{\bibfnamefont{G.}~\bibnamefont{Shirane}},
  \bibinfo{author}{\bibfnamefont{R.~J.} \bibnamefont{Birgeneau}},
  \bibinfo{author}{\bibfnamefont{P.~M.} \bibnamefont{Richards}},
  \bibnamefont{and} \bibinfo{author}{\bibfnamefont{S.~L.} \bibnamefont{Holt}},
  \bibinfo{journal}{Phys. Rev. Lett.} \textbf{\bibinfo{volume}{32}},
  \bibinfo{pages}{170} (\bibinfo{year}{1974}).

\bibitem[{\citenamefont{Clarke and Harrison}(1992)}]{Clarke92b}
\bibinfo{author}{\bibfnamefont{S.}~\bibnamefont{Clarke}} \bibnamefont{and}
  \bibinfo{author}{\bibfnamefont{A.}~\bibnamefont{Harrison}},
  \bibinfo{journal}{J. Phys. Cond. Mat.} \textbf{\bibinfo{volume}{4}},
  \bibinfo{pages}{6217} (\bibinfo{year}{1992}).

\bibitem[{\citenamefont{Yamagata and Abe}(1983)}]{Yamagata83}
\bibinfo{author}{\bibfnamefont{K.}~\bibnamefont{Yamagata}} \bibnamefont{and}
  \bibinfo{author}{\bibfnamefont{H.}~\bibnamefont{Abe}}, \bibinfo{journal}{J.
  Mag. Mag. Mat.} \textbf{\bibinfo{volume}{31-34}}, \bibinfo{pages}{1179}
  (\bibinfo{year}{1983}).

\bibitem[{\citenamefont{Koyama et~al.}(1987)\citenamefont{Koyama, Nobumasa, and
  Matsura}}]{Koyama87}
\bibinfo{author}{\bibfnamefont{K.}~\bibnamefont{Koyama}},
  \bibinfo{author}{\bibfnamefont{H.}~\bibnamefont{Nobumasa}}, \bibnamefont{and}
  \bibinfo{author}{\bibfnamefont{M.}~\bibnamefont{Matsura}},
  \bibinfo{journal}{J. Phys. Soc Jpn.} \textbf{\bibinfo{volume}{56}},
  \bibinfo{pages}{1553} (\bibinfo{year}{1987}).

\bibitem[{\citenamefont{Hayden et~al.}(1991)\citenamefont{Hayden, Aeppli,
  Osborn, Taylor, Perring, Cheong, and Fist}}]{Hayden91}
\bibinfo{author}{\bibfnamefont{S.~M.} \bibnamefont{Hayden}},
  \bibinfo{author}{\bibfnamefont{G.}~\bibnamefont{Aeppli}},
  \bibinfo{author}{\bibfnamefont{R.}~\bibnamefont{Osborn}},
  \bibinfo{author}{\bibfnamefont{A.~D.} \bibnamefont{Taylor}},
  \bibinfo{author}{\bibfnamefont{T.~G.} \bibnamefont{Perring}},
  \bibinfo{author}{\bibfnamefont{S.~W.} \bibnamefont{Cheong}},
  \bibnamefont{and} \bibinfo{author}{\bibfnamefont{Z.}~\bibnamefont{Fist}},
  \bibinfo{journal}{Phys. Rev. Lett} \textbf{\bibinfo{volume}{67}},
  \bibinfo{pages}{3622} (\bibinfo{year}{1991}).

\bibitem[{\citenamefont{MacLaughlin et~al.}(1994)\citenamefont{MacLaughlin,
  Vithayathil, and et~al.}}]{Maclaughlin94}
\bibinfo{author}{\bibfnamefont{D.~E.} \bibnamefont{MacLaughlin}},
  \bibinfo{author}{\bibfnamefont{J.~P.} \bibnamefont{Vithayathil}},
  \bibnamefont{and} \bibinfo{author}{\bibnamefont{et~al.}},
  \bibinfo{journal}{Phys. Rev. Lett.} \textbf{\bibinfo{volume}{72}},
  \bibinfo{pages}{760} (\bibinfo{year}{1994}).

\bibitem[{\citenamefont{Chakravarty et~al.}(1991)\citenamefont{Chakravarty,
  Gelfand, Kopietz, Orbach, and Wollensak}}]{Chakravarty91}
\bibinfo{author}{\bibfnamefont{S.}~\bibnamefont{Chakravarty}},
  \bibinfo{author}{\bibfnamefont{M.}~\bibnamefont{Gelfand}},
  \bibinfo{author}{\bibfnamefont{P.}~\bibnamefont{Kopietz}},
  \bibinfo{author}{\bibfnamefont{R.}~\bibnamefont{Orbach}}, \bibnamefont{and}
  \bibinfo{author}{\bibfnamefont{M.}~\bibnamefont{Wollensak}},
  \bibinfo{journal}{Phys. Rev. B} \textbf{\bibinfo{volume}{43}},
  \bibinfo{pages}{2796} (\bibinfo{year}{1991}).

\bibitem[{\citenamefont{Imai et~al.}(1993)\citenamefont{Imai, Slichter,
  Yoshimura, Katoh, and Kosuge}}]{Imai93}
\bibinfo{author}{\bibfnamefont{T.}~\bibnamefont{Imai}},
  \bibinfo{author}{\bibfnamefont{C.}~\bibnamefont{Slichter}},
  \bibinfo{author}{\bibfnamefont{K.}~\bibnamefont{Yoshimura}},
  \bibinfo{author}{\bibfnamefont{M.}~\bibnamefont{Katoh}}, \bibnamefont{and}
  \bibinfo{author}{\bibfnamefont{K.}~\bibnamefont{Kosuge}},
  \bibinfo{journal}{Phys. Rev. Lett.} \textbf{\bibinfo{volume}{71}},
  \bibinfo{pages}{1254,} (\bibinfo{year}{1993}).

\bibitem[{\citenamefont{Castner and Seehra}(1993)}]{Castner93}
\bibinfo{author}{\bibfnamefont{T.~G.} \bibnamefont{Castner}} \bibnamefont{and}
  \bibinfo{author}{\bibfnamefont{M.~S.} \bibnamefont{Seehra}},
  \bibinfo{journal}{Phys. Rev. B} \textbf{\bibinfo{volume}{47}},
  \bibinfo{pages}{578} (\bibinfo{year}{1993}).

\bibitem[{\citenamefont{Abrahams}(1962)}]{Abrahams62}
\bibinfo{author}{\bibfnamefont{S.~C.} \bibnamefont{Abrahams}},
  \bibinfo{journal}{J. Chem. Phys.} \textbf{\bibinfo{volume}{36}},
  \bibinfo{pages}{56} (\bibinfo{year}{1962}).

\bibitem[{\citenamefont{Woodward et~al.}(2001)\citenamefont{Woodward, Landee,
  Giantsidis, Turnbull, and Richardson}}]{5bap-art}
\bibinfo{author}{\bibfnamefont{F.~M.} \bibnamefont{Woodward}},
  \bibinfo{author}{\bibfnamefont{C.~P.} \bibnamefont{Landee}},
  \bibinfo{author}{\bibfnamefont{J.}~\bibnamefont{Giantsidis}},
  \bibinfo{author}{\bibfnamefont{M.~M.} \bibnamefont{Turnbull}},
  \bibnamefont{and}
  \bibinfo{author}{\bibfnamefont{C.}~\bibnamefont{Richardson}},
  \bibinfo{journal}{Inorganica Chemica Acta} \textbf{\bibinfo{volume}{Accepted
  for publication Jun 2001}} (\bibinfo{year}{2001}).

\bibitem[{\citenamefont{Landee et~al.}(2001)\citenamefont{Landee, Turnbull,
  Galeriu, Giantsidis, and Woodward}}]{IAP}
\bibinfo{author}{\bibfnamefont{C.}~\bibnamefont{Landee}},
  \bibinfo{author}{\bibfnamefont{M.}~\bibnamefont{Turnbull}},
  \bibinfo{author}{\bibfnamefont{C.}~\bibnamefont{Galeriu}},
  \bibinfo{author}{\bibfnamefont{J.}~\bibnamefont{Giantsidis}},
  \bibnamefont{and} \bibinfo{author}{\bibfnamefont{F.}~\bibnamefont{Woodward}},
  \bibinfo{journal}{Phys. Rev. B. Rapid Comm.} \textbf{\bibinfo{volume}{63}},
  \bibinfo{pages}{100402} (\bibinfo{year}{2001}).

\bibitem[{\citenamefont{Albrecht et~al.}(1997)\citenamefont{Albrecht, Landee,
  Slanic, and Turnbull}}]{pz-bf4}
\bibinfo{author}{\bibfnamefont{A.~S.} \bibnamefont{Albrecht}},
  \bibinfo{author}{\bibfnamefont{C.~P.} \bibnamefont{Landee}},
  \bibinfo{author}{\bibfnamefont{Z.}~\bibnamefont{Slanic}}, \bibnamefont{and}
  \bibinfo{author}{\bibfnamefont{M.~M.} \bibnamefont{Turnbull}},
  \bibinfo{journal}{Mol. Cryst. Liq. Cryst.} \textbf{\bibinfo{volume}{305}},
  \bibinfo{pages}{333} (\bibinfo{year}{1997}).

\bibitem[{\citenamefont{Endoh et~al.}(1988)\citenamefont{Endoh, Yamada,
  Birgeneau, and et~al.}}]{Endoh88}
\bibinfo{author}{\bibfnamefont{Y.}~\bibnamefont{Endoh}},
  \bibinfo{author}{\bibfnamefont{K.}~\bibnamefont{Yamada}},
  \bibinfo{author}{\bibfnamefont{R.~J.} \bibnamefont{Birgeneau}},
  \bibnamefont{and} \bibinfo{author}{\bibnamefont{et~al.}},
  \bibinfo{journal}{Phys. Rev. B} \textbf{\bibinfo{volume}{37}},
  \bibinfo{pages}{7443} (\bibinfo{year}{1988}).

\bibitem[{\citenamefont{Coldea et~al.}(2001)\citenamefont{Coldea, Hayden,
  Aeppli, Perring, Frost, Mason, Cheong, and Fisk}}]{coldea01}
\bibinfo{author}{\bibfnamefont{R.}~\bibnamefont{Coldea}},
  \bibinfo{author}{\bibfnamefont{S.~M.} \bibnamefont{Hayden}},
  \bibinfo{author}{\bibfnamefont{G.}~\bibnamefont{Aeppli}},
  \bibinfo{author}{\bibfnamefont{T.~G.} \bibnamefont{Perring}},
  \bibinfo{author}{\bibfnamefont{C.~D.} \bibnamefont{Frost}},
  \bibinfo{author}{\bibfnamefont{T.~E.} \bibnamefont{Mason}},
  \bibinfo{author}{\bibfnamefont{S.-W.} \bibnamefont{Cheong}},
  \bibnamefont{and} \bibinfo{author}{\bibfnamefont{Z.}~\bibnamefont{Fisk}},
  \bibinfo{journal}{Phys. Rev. Lett.} \textbf{\bibinfo{volume}{86}},
  \bibinfo{pages}{5377} (\bibinfo{year}{2001}).

\bibitem[{\citenamefont{Rønnow et~al.}(2001)\citenamefont{Rønnow, McMorrow,
  Coldea, Harrison, Youngson, Perring, Aeppli, Syljuåsen, Lefmann, and
  Rischel}}]{ronnow01}
\bibinfo{author}{\bibfnamefont{H.~M.} \bibnamefont{Rønnow}},
  \bibinfo{author}{\bibfnamefont{D.~F.} \bibnamefont{McMorrow}},
  \bibinfo{author}{\bibfnamefont{R.}~\bibnamefont{Coldea}},
  \bibinfo{author}{\bibfnamefont{A.}~\bibnamefont{Harrison}},
  \bibinfo{author}{\bibfnamefont{I.~D.} \bibnamefont{Youngson}},
  \bibinfo{author}{\bibfnamefont{T.~G.} \bibnamefont{Perring}},
  \bibinfo{author}{\bibfnamefont{G.}~\bibnamefont{Aeppli}},
  \bibinfo{author}{\bibfnamefont{O.}~\bibnamefont{Syljuåsen}},
  \bibinfo{author}{\bibfnamefont{K.}~\bibnamefont{Lefmann}}, \bibnamefont{and}
  \bibinfo{author}{\bibfnamefont{C.}~\bibnamefont{Rischel}},
  \bibinfo{journal}{Phys. Rev. Lett.} \textbf{\bibinfo{volume}{87}},
  \bibinfo{pages}{20737} (\bibinfo{year}{2001}).

\bibitem[{\citenamefont{Algra et~al.}(1978)\citenamefont{Algra, de~Jongh, and
  Carlin}}]{Algra78}
\bibinfo{author}{\bibfnamefont{H.}~\bibnamefont{Algra}},
  \bibinfo{author}{\bibfnamefont{L.}~\bibnamefont{de~Jongh}}, \bibnamefont{and}
  \bibinfo{author}{\bibfnamefont{R.}~\bibnamefont{Carlin}},
  \bibinfo{journal}{Physica} \textbf{\bibinfo{volume}{93B}},
  \bibinfo{pages}{24} (\bibinfo{year}{1978}).

\bibitem[{\citenamefont{Lumsden et~al.}(2001)\citenamefont{Lumsden, Sales,
  Mandrus, Nagler, and Thompson}}]{lumsden}
\bibinfo{author}{\bibfnamefont{M.~D.} \bibnamefont{Lumsden}},
  \bibinfo{author}{\bibfnamefont{B.~C.} \bibnamefont{Sales}},
  \bibinfo{author}{\bibfnamefont{D.}~\bibnamefont{Mandrus}},
  \bibinfo{author}{\bibfnamefont{S.~E.} \bibnamefont{Nagler}},
  \bibnamefont{and} \bibinfo{author}{\bibfnamefont{J.~R.}
  \bibnamefont{Thompson}}, \bibinfo{journal}{Phys. Rev. Lett.}
  \textbf{\bibinfo{volume}{86}}, \bibinfo{pages}{159} (\bibinfo{year}{2001}).

\end{thebibliography}
\clearpage
\begin{table}
\caption{Crystal data and structure refinement for (5CAP)$_{2}$CuBr$_{4}$.}

{
\begin{tabular}{ll}
\hline
\hline
Space group C2/c & Formula weight 642.32\\
\hline
a = 13.050(5) \AA &T = -130(2)${ }^{\circ}$C \\
b = 8.769(3) \AA & $\lambda$ = 0.71073 \AA\\
c = 15.810(5) \AA &$\rho_{Cal} = 2.365$ g cm$^{-3}$\\
$\beta$ = 94.31(3)$^\circ$ & $\mu$ = 10.362 mm$^{-1}$\\
& Transmission coefficient = 0.12431--0.47808\\
V = {1804.1(11) \AA$^{3}$} & R(F$_0$) = 0.0476 (= 0.0679, all reflection)\\
Z = 4 & R$_w$(F$_0$) = 0.1053 (= 0.1153, all reflection)\\
\hline
\hline
\end{tabular}
}
\label{crys}
\end{table}

\begin{table}
\caption{Atomic coordinates ( x $10^4$) and equivalent isotropic 
displacement
parameters  (\AA$^2$ x $10^3$)
for(5CAP)$_{2}$CuBr$_{4}$. U(EC) is defined as one third of the trace of 
the orthogonality $U_{ij}$ tensor.}
\label{table2}
\begin{tabular}{lllll}
\hline
\hline
&x&y&z&U(EC)\\
\colrule
Br(1) & 1368(1) &886(1) & 3468(1) & 21(1)\\
Br(2) & -1042(1) & -1079(1) & 3512(1) & 22(1) \\
Cu & 0000 & 47(1) & 2500 & 17(1)\\
Cl & 751(2) & -4116(3) & 6704(1) & 32(1)\\
N(1) & 1269(5) & -2430(7) & 4476(4) & 21(1)\\
C(1) & 1607(6) & -3618(8) & 4016(5) & 17(2)\\
C(2) & 1653(6) & -5069(8) & 4412(5) & 20(2)\\
N(2) & 1882(5) & -3359(7) & 3230(4) & 26(2)\\
C(3) & 1383(6) & -5210(8) & 5222(5) & 20(2)\\
C(4) & 1063(5) & -3941(9) & 5660(5) & 20(2)\\
C(5) & 983(6) & -2573(8) & 5281(5) & 19(2)\\
\hline
\hline
\end{tabular}
\end{table}

\begin{table}
\caption{Selected bond distances (\AA) and angles ($^o$) for 
(5CAP)$_{2}$CuBr$_{4}$.}
\label{table3}
\begin{tabular}{lllll}
\hline
\hline
&&Bond distances (\AA)&&\\
\hline
Br(1)---Cu & 2.3792(12) && Br(2)---Cu& 2.3897(11)\\
Cl---C(4) & 1.738(8) && N(1)---C(5) & 1.359(10)\\
N(1)---C(1) & 1.362(9) && C(1)---N(2) & 1.338(10)\\
C(1)---C(2) & 1.417(10) && C(2)---C(3) & 1.360(11)\\
C(3)---C(4) & 1.390(10) && C(4)---C(5) & 1.342(11)\\
\hline
\\
\hline
&& Bond angles (DEC)&&\\
\hline
Br(1)'---Cu---Br(1) & 143.95(7) && Br(1)---Cu---Br(2) & 97.71(4)\\
Br(1)---Cu---Br(2)' & 96.97(4) && Br(2)'---Cu---Br(2) & 131.20(7)\\
C(5)---N(1)---C(1) & 123.5(6) && N(2)---C(1)---N(1) & 119.1(7)\\
N(2)---C(1)---C(2) & 123.8(7) && N(1)---C(1)---C(2) & 117.1(6)\\
C(3)---C(2)---C(1) & 119.5(7) && C(2)---C(3)---C(4) & 120.3(7)\\
C(5)---C(4)---C(3) & 120.7(7) && C(5)---C(4)---Cl & 119.1(6)\\
C(3)---C(4)---Cl & 120.2(6) && C(4)---C(5)---N(1) & 118.9(7)\\
\hline
\hline
\end{tabular}
\end{table}

\begin{table}
\caption{Examples of S=1/2 2D Heisenberg
Antiferromagnets.}
\label{2ds}
\begin{tabular}{llllll}
\hline
\hline
Compound & J/k (K) & T$_N$ (K) & T$_N$/J & Comments & Ref.\\
\hline
La$_{2}$CuO$_{4}$ & $\approx$1500 & 310 & 0.21 & Slight hidden canting &
 \cite{Endoh88,coldea01}\\
Sr$_{2}$CuO$_{2}$Cl$_{2}$ & $\approx$1450 & 251 & 0.17 & no canting & 
\cite{Greven95}\\
Cu(COO)$_{2}{\cdot}$4H$_{2}$O & 70 & 16.5 & 0.24 & CuF4H, 
canting & \cite{Clarke92b,Ronnow99,ronnow01}\\
Cu(COO)$_{2}{\cdot}$2CO(NH$_{2})_{2}{\cdot}$2H$_{2}$O & 70 & 
15.5 &
0.23 & CuFUH, canting & \cite{Yamagata83}\\
CuF$_{2}{\cdot}$2H$_{2}$0 & 26 & 10.9 & 0.42 & Higher 3d interactions & 
\cite{Koyama87}\\
$\mathrm{\left[Cu(PyNO)_{6}\right]\left[BF_{4}\right]_{2}}$ & 4.4 & 
0.62 & 0.28 & Structural transition & \cite{Algra78}\\
K$_{2}$V$_{3}$O$_{8}$& 12.6 & 4.0 & 0.32 & spin canting & \cite{lumsden}\\
\hline
(5CAP)$_{2}$CuBr$_{4}$ & 8.5 & 5.08 & 0.60 & strong 3d interactions & 
this work,\cite{Sorai96}\\
(5MAP)$_{2}$CuBr$_{4}$ & 6.5 & 3.8 & 0.58 & strong 3d interactions & 
this work\\
(5CAP)$_{2}$CuCl$_{4}$ & 1.14 & 0.74 & 0.64 & strong 3d interactions & 
\cite{Hammar97}\\
(5MAP)$_{2}$CuCl$_{4}$ & 0.76 & 0.44 & 0.57 & strong 3d interactions & 
\cite{Hammar97}\\
\hline
\hline
\end{tabular}
\end{table}

\begin{table}
\caption{2D QHAF model polynomial coefficients}
\label{coef}
\begin{tabular}{lll}
\hline
\hline
n & $a_{n}$ & $b_{n}$ \\
\hline
1 & 0.998586 & -1.84279 \\

2 & -1.28534 & 1.14141 \\

3 & 0.656313 & -0.704192 \\

4 & 0.235862 & -0.189044 \\

5 & 0.277527 & -0.277545 \\
\hline
\hline
\end{tabular}
\end{table}


\end{document}